\documentclass[12]{article}
\usepackage{amsmath,color,xcolor}
\usepackage{graphics,graphicx,amsbsy,amssymb}

\newcommand{\PP}{{\boldmath \mbox{$P$}}}
\newcommand{\JJ}{{\boldmath \mbox{$J$}}}

\newcommand{\aaa}{{\boldmath \mbox{$a$}}}

\newcommand{\uu}{{\boldmath \mbox{$u$}}}
\newcommand{\bb}{{\boldmath \mbox{$b$}}}

\newcommand{\xx}{{\boldmath \mbox{$x$}}}

\newcommand{\rr}{{\boldmath \mbox{$r$}}}

\newcommand{\qq}{{\boldmath \mbox{$q$}}}
\newcommand{\cc}{{\boldmath \mbox{$c$}}}

\newcommand{\pp}{{\boldmath \mbox{$p$}}}
\newcommand{\vv}{{\boldmath \mbox{$v$}}}

\newcommand{\ddelta}{{\boldmath\mbox{$\delta$}}}

\newcommand{\ssigma}{{\boldmath\mbox{$\sigma$}}}

\newlength{\defbaselineskip}
\newcommand{\setlinespacing}[1]%
           {\setlength{\baselineskip}{#1 \defbaselineskip}}

\begin{document}

\title{\textbf{ Hamiltonian and Godunov Structures of the Grad Hierarchy}}
\author{Miroslav Grmela$^{1}$\footnote{corresponding author: e-mail:
miroslav.grmela@polymtl.ca},$\;$ Liu Hong$^{2}$, David Jou$^{3}$,\\ Georgy Lebon$^{4}$, Michal Pavelka $^{1,5}$\vspace {0.5cm}\\
$^1$ \'{E}cole Polytechnique de Montr\'{e}al,\\
  C.P.6079 suc. Centre-ville,
 Montr\'{e}al, H3C 3A7,  Qu\'{e}bec, Canada \\
$^2$ Zhou Pei-Yuan Center for Applied Mathematics,\\ Tsinghua University, Beijing, China \\
$^3$ Department de Fisica, Universitat Autonoma de Barcelona,\\ 08193 Bellaterra, Catalonia, Spain \\
$^4$ Thermodynamique des Ph\'{e}nom\`{e}nes Irr\'{e}versibles, Universit\'{e} de Li\`{e}ge\\
Quartier Agora, All\'{e}e 6 Ao\^{u}t, 19, 4000 Li\`{e}ge, Belgique \\
$^5$ Mathematical Institute, Faculty of Mathematics, Charles University,\\ Prague,
 Sokolovsk\'{a} 83, 18675 Prague, Czech Republic
 }
 \date{}
\maketitle

\begin{abstract}

The time evolution governed by the Boltzmann kinetic equation is compatible with mechanics and thermodynamics. The former compatibility is mathematically expressed in the Hamiltonian and  Godunov structures, the latter  in the structure of gradient dynamics guaranteeing the growth of entropy and consequently the approach  to equilibrium.
We carry all three structures  to the Grad reformulation  of the Boltzmann equation (to the Grad hierarchy). First, we recognize the structures in the infinite Grad hierarchy and then in several examples of  finite hierarchies representing  extended hydrodynamic equations. In the context of Grad's hierarchies we also investigate relations between  Hamiltonian and Godunov structures.

\end{abstract}

\section{Introduction}

Behavior of complex fluids, i.e. fluids  involving an internal structure (either induced by external forces - as e.g. structures in turbulent flows - or structures   of suspended particles or macromolecules),  is not well described by classical fluid mechanics. A more microscopic theory (i.e. a theory involving more details) is needed. The Boltzmann kinetic theory (in which the one particle distribution function $f(\rr,\vv)$  is  playing the role of the state variable; $\rr$ is the position vector and $\vv$ momentum of one particle)
appears to be a natural candidate for such theory.  Of course, the Boltzmann theory addresses only the behavior of  ideal gases but it can still serve as a useful starting point for developing an extended fluid mechanics. Indeed, Grad's reformulation of the Boltzmann kinetic equation  (consisting of replacing $f(\rr,\vv)$ with an infinite number of fields $(c^{(0)}(\rr),...,c^{(\infty)}(\rr))$)
has played a significant  role in  such research.
Among the important  questions that arise  in the investigations of the passage \textit{Boltzmann theory} $\rightarrow$ \textit{extended fluid mechanics}  we mention  the following:
How does the Boltzmann kinetic theory reduce to mesoscopic fluid mechanics theories involving less microscopic details?
What are the properties of solutions of the Boltzmann kinetic equation that do and that do not pass to the mesoscopic fluid mechanics  theories and eventually to equilibrium thermodynamics? What are the new  properties of solutions  that emerge in  reduced theories?

To answer this type of questions, we proceed in the following two steps.
First, in Section \ref{BE}, we recognize  in the Boltzmann kinetic equation a structure of physical significance and  then we keep it in Grad's reformulation and in reductions. By a  "structure of physical significance" we mean a structure  of the time evolution equations that guarantees that their solutions  agree with  results of certain basic experimental observations. Our focus on structures is a mathematical representation of our focus on physical understanding and experimental validation of reduced theories.
The structure that we require to be passed from the Boltzmann equation to all its reformulations and eventually to all its reductions is the following (see more details in Section \ref{BE}): (i) the vector field of  the Boltzmann equation is a sum of time reversible part (the free flow term) and the time irreversible part (the Boltzmann collision term), (ii) the time reversible part represents Hamiltonian dynamics (i.e. it has the form $LE_f$, where $E_f=\frac{\partial E}{\partial f(\rr,\vv)}$ is the gradient of the energy $E(f)$, $f(\rr,\vv)$ is the one particle distribution function, and $L$ is a Poisson bivector transforming a co-vector into a vector), (iii) the time reversible part possesses also the Godunov structure, (iv) the entropy remains unchanged during the time reversible time evolution, (v)
the time irreversible part represents  a generalized gradient dynamics in which the energy $E(f)$ remains unchanged and the entropy growths. The structure (ii) is physically significant because it expresses the compatibility of the Boltzmann kinetic theory with mechanics (the classical mechanics of particles inherited in the reduced description that uses $f(\rr,\vv)$ as state variable). The structures (iv) and (v) express the compatibility of the Boltzmann theory with thermodynamics (i.e. an agreement between theoretical predictions and experimentally observed approach to equilibrium at which the classical equilibrium thermodynamics is found to describe well the observed behavior).

Having identifies the mathematical structure of importance, we then  make, in the second step,  Grad's reformulation separately for every structure.
Reduced equations (i.e. governing equations of extended fluid mechanics) are  constructed as particular realizations of the Grad form of the structures. In fact, we are constructing in this way both reduced equations (i.e. equations governing the time evolution of $(c^{(0)}(\rr),...,c^{(\infty)}(\rr))$ ) and reducing equations (i.e. equations governing the time evolution of $(c^{(N+1)}(\rr),...,c^{(\infty)}(\rr))$).

This two-step approach to reductions distinguishes the investigation presented in this paper from the investigations reported in
\cite{MR},  \cite{Jou},\cite{Rugg}, \cite{Struch}. After developing our approach we discuss briefly its relation to other approaches in Section \ref{Compoth}.

\section{Hamiltonian Boltzmann  equation}\label{BE}

Let $f(\rr,\vv)$  be  one particle distribution function, $\rr$ denotes the position vector and $\vv$ momentum of one particle. We are taking in this paper the viewpoint of the time evolution that is common in the dynamical system theory. We regard $f(\rr,\vv)$  as an element of the state space. The time evolution  of $f(\rr,\vv)$  (i.e. $f(\rr,\vv,t)$)  is generated by a vector field, that is the right hand side of the time evolution equation $\frac{\partial f}{\partial t}=...$). We therefore include the time $t$ into the set of variables on which $f$ depends only if we refer to $f$ that is a solution of a particular time evolution equation.

The vector field generating  the time evolution in Boltzmann's kinetic theory is a
sum of reversible and irreversible parts \cite{PKG}. The former changes the sign under the transformation $\vv\rightarrow -\vv$ and the latter remains invariant. In  discussions of the Boltzmann equation as well as in discussions of all its reformulations we  shall always consider first the reversible part and then the irreversible part.
The main objective of Sections \ref{GodB}  and \ref{HamB}  is to define the Hamiltonian and Godunov structures and to prepare the setting for investigating Grad's hierarchy.

\subsection{Classical formulation of the reversible  Boltzmann equation}\label{CBE}

The reversible part of the Boltzmann equation is given by
\begin{equation}\label{BErev}
\left(\frac{\partial f(\rr,\vv)}{\partial t}\right)_{rev}=-\frac{\partial}{\partial \rr}\left(\frac{\vv}{m} f(\rr,\vv)\right)
\end{equation}
where $m$ is mass of the particle. We directly verify that the right hand side of (\ref{BErev}) changes indeed its sign if $\vv\rightarrow -\vv$. Equation (\ref{BErev}) is a continuity (Liouville) equation corresponding to the one particle motion governed by
\begin{equation}\label{BEpH}
\left(\begin{array}{cc}\dot{\rr}\\ \dot{\vv}\end{array}\right)=L^{(p)}\left(\begin{array}{cc}E^{(p)}_{\rr}\\E^{(p)}_{\vv}\end{array}\right)
\end{equation}
where the dot denotes time derivative, $E^{(p)}(\rr,\pp)=\frac{\vv^2}{2m}$ is the kinetic energy of one particle, $E^{(p)}_{\rr}=\frac{\partial E^{(p)}}{\partial \rr}$,  $E^{(p)}_{\vv}=\frac{\partial E^{(p)}}{\partial \vv}$, and $L^{(p)}=\left(\begin{array}{cc}0&1\\-1&0\end{array}\right)$.

Equation (\ref{BErev}) is a simple equation that can be exactly solved. Its solution is
\begin{equation}\label{solBE}
(f_0(\rr,\vv))_{t=t_0}\rightarrow
f_0(\rr-\frac{\vv}{m}t,\vv)
\end{equation}

For the investigations below,
more important than the exact solutions (\ref{solBE}) are the following two qualitative properties of solutions to (\ref{BErev}).   First, it is the conservation of energy
$E(f)$ given by
\begin{equation}\label{Energ}
E(f)=\int d\rr\int d\vv e(f,\rr,\vv)
\end{equation}
with
\begin{equation}\label{Ener}
e(f,\rr,\vv)=\frac{\vv^2}{2m}f(\rr,\vv)
\end{equation}
We easily see that
\begin{equation}\label{energ}
\left(\frac{\partial e}{\partial t}\right)_{rev}=-\frac{\partial \left(e \frac{\vv}{m}\right)}{\partial\rr}
\end{equation}
and thus
\begin{equation}\label{Edot}
\dot{E}=0
\end{equation}
provided the integral over the boundaries equals zero. In this paper we limit our investigation to externally unforced systems. This means, in particular, that the boundary conditions guarantee that all the integrals over boundaries that arise in integrations over $\rr$ equal zero. The property (\ref{Edot}) can, of course, be anticipated from the conservation of the particle energy $E^{(p)}(\rr,\vv)$ in the time evolution governed by (\ref{BEpH}).

The second important qualitative property of solutions to (\ref{BErev}) is not, on the other hand, seen in the particle mechanics (\ref{BEpH}). It emerges only in its Liouville formulation (\ref{BErev}).
 Let $S(f)$, called hereafter entropy, be given by
\begin{eqnarray}\label{Cas}
&&S(f)=\int d\rr\int d\vv \eta(f(\rr,\vv))\nonumber \\
&&\eta:\mathbb{R}\rightarrow \mathbb{R}\,\, is\,\, a\,\, sufficiently\,\, regular\,\, function
\end{eqnarray}
We easily verify that
\begin{equation}\label{eta}
\left(\frac{\partial \eta}{\partial t}\right)_{rev}=-\frac{\partial \left(\eta\frac{\vv}{m}\right)}{\partial\rr}
\end{equation}
and thus
\begin{equation}\label{Sdot}
\dot{S}=0
\end{equation}
Specifically, $S(f)$ becomes the Boltzmann entropy if we choose
\begin{equation}\label{Bent}
\eta(f)=-k_Bf\ln f
\end{equation}
where $k_B$ is the Boltzmann constant. The qualitative property (\ref{eta}) of solutions to (\ref{BErev}) does not indeed originate in mechanics (\ref{BEpH}). We shall see that, from the physical point of view, it relates mechanics to thermodynamics. The  choice of $\eta(f)$ is shown in  Section \ref{IrrBE} to be closely related to the choice of the irreversible part of the right hand side of the Boltzmann equation and thus to the approach to thermodynamic equilibrium.

In view of the importance of the two properties (\ref{energ}) and (\ref{eta}) in the investigation of the passage from kinetic theory to hydrodynamics, we shall,
in the rest of this section,  reformulate (\ref{BErev}) into   forms that manifestly display them. First, we cast (\ref{BErev}) into the Godunov form displaying (\ref{eta}) and then into the Hamiltonian form that manifestly displays both (\ref{energ}) and (\ref{eta}).

Before entering the reformulations of (\ref{BErev}) we recall an important result proven in \cite{Villani}. Solutions to (\ref{BErev}) converge weakly in long time to spatially homogeneous distribution  that is a space average of the initial distribution function.

\subsection{Godunov formulation of the reversible Boltzmann equation}\label{GodB}

Following Godunov's   investigations  of partial differential equations  \cite{Godunov}, \cite{FL}, \cite{WAY}, we turn now our attention to a special class of partial differential equations that have the form of local conservation laws (i.e. the time derivative equals divergence of a flux). For example, Eq.(\ref{BErev}) is a local conservation law. Godunov has noted that if there exists a sufficiently regular and convex function of the fields playing the role of the state variables whose time evolution is also governed by a local conservation law (called a companion local conservation law) then this property has important physical and mathematical consequences. Godunov moreover identified a structure in the system of local conservation laws that makes the existence of the companion conservation law manifestly visible. We call this structure a Godunov structure. We shall demonstrate it on the example of  Eq.(\ref{BErev}).

We begin by choosing $\eta:\mathbb{R}\rightarrow \mathbb{R}$ that is sufficiently regular and convex (or concave). We define
\begin{equation}\label{conj}
f^*=\eta_f
\end{equation}
and call it a conjugate of $f$ with respect to $\eta$ or simply a conjugate of $f$ (with the understanding that the function $\eta$ is fixed). The Legendre transformation $\eta^*(f^*)$ of $\eta(f)$ is given by $\eta^*(f^*)=-\eta(f(f^*))+f^*f(f^*)$, where $f(f^*)$ is a solution of $(-\eta(f)+f^*f)_f=0$.
We note that $f$ is the conjugate of $\eta^*(f^*)$ (i.e. $f=\eta^*_{f^*})$ and also that $\eta(f)=-\eta^*(f^*(f))+ff^*(f)$, where $f^*(f)$ is a solution to $(-\eta^*(f^*)+ff^*)_{f^*}=0$. We can easily verify that (\ref{BErev}) is equivalent to
\begin{eqnarray}\label{BEG}
&&\frac{\partial \eta^*_{f^*}}{\partial t}=-\frac{\partial}{\partial \rr}\JJ^{(r)*}_{f^*}+\frac{\partial}{\partial\vv}\JJ^{(v)*}_{f^*}\nonumber \\
&&\JJ^{(r)*}(f^*)=\frac{\partial E^{(p)}}{\partial\vv}\eta^*(f^*);\,\JJ^{(v)}(f^*)=\frac{\partial E^{(p)}}{\partial\rr}\eta^*(f^*)
\end{eqnarray}
provided $E^{(p)}(\rr,\vv)=\frac{\vv^2}{2m}$. We interpret  $\JJ^*=(\JJ^{(r)*},\JJ^{(v)*})$ as flux of $f$. We note that the essence of the reformulation (\ref{BEG}) is to regard the distribution function $f$ in (\ref{BErev}) as a conjugate of conjugate (i.e. $f=\eta^*_{f^*}$) and realize that the equation governing the time  evolution  of $f$ expressed in this way is an equation of the same type (i.e. a local conservation law) as the equation  governing the time evolution of $f$. This means that if we succeed to cast (\ref{BErev}) (and in general a system of local conservations laws) into the Godunov form ((\ref{BEG}) in the case of the Boltzmann equation (\ref{BErev})) then the existence of the companion local conservation law (see the Godunov Property 1 below) is immediately visible.
We also note that the flux $\JJ^*(f^*)$ arising in the vector field is a function of the conjugate $f^*$ of $f$. The same property is displayed also in the Hamiltonian formulation (see (\ref{BEH}) below)  but with a different conjugacy relation. In (\ref{BEG}) the conjugate state variables are obtained by deriving the Casimir $S(f)$  (see (\ref{etaH}) below) with respect to $f$ (i.e.$f^*=\eta(f)_f$) and in (\ref{BEH}) by deriving the energy $E(f)$ with respect to $f$ (i.e.$f^*=E(f)_f$).

Equation  (\ref{BEG}) has two consequences:

\textit{\textbf{Godunov Property 1}}

By multiplying (\ref{BEG}) by $f^*$ we obtain
\begin{eqnarray}\label{BEc}
\frac{\partial}{\partial t}\left(-\eta^*+<f^*,\eta^*_{f^*}>\right)&=&-\frac{\partial}{\partial\rr}\left(<f^*,\JJ^{(r)*}_{f^*}>-\JJ^{(r)*}\right)\nonumber \\ &&+\frac{\partial}{\partial\vv}\left(<f^*,\JJ^{(v)*}_{f^*}>-\JJ^{(v)*}\right)
\end{eqnarray}
If we take into account $\eta(f)=-\eta^*(f^*(f))+ff^*(f)$ (and provided $E^{(p)}(\rr,\vv)=\frac{\vv^2}{2m}$) then (\ref{BEc}) is equivalent to (\ref{eta}) since $f^*\JJ^{(r)*}_{f^*}-\JJ^{(r)*}=
\frac{\partial E^p}{\partial\vv}\eta(f)$ and
$f^*\JJ^{(v)*}_{f^*}-\JJ^{(v)*}=\frac{\partial E^p}{\partial\rr}\eta(f)$ if $\JJ^{(r)*}$ and $\JJ^{(v)*}$ are given by the second line in (\ref{BEG}). From this conclusion we see that
the Godunov formulation (\ref{BEG})  manifestly displays the local conservation equation (\ref{eta}) governing the time evolution of the Casimir density $\eta(f)$.

\textit{\textbf{Godunov Property 2}}

From (\ref{BEG}) we immediately obtain
\begin{equation}\label{GG}
\eta^*_{f^*f^*}\frac{\partial f^*}{\partial t}=-\JJ^{(r)*}_{f^*f^*}\frac{\partial f^*}{\partial\rr}+\JJ^{(v)*}_{f^*f^*}\frac{\partial f^*}{\partial\vv}
\end{equation}
In the context of Grad's hierarchy this consequence will imply that the Cauchy problem for the partial differential equation appearing in the reversible part of Grad's hierarchy is well posed.

Summing up, the Godunov formulation  (\ref{BEc}) of the reversible Boltzmann equation (\ref{BErev}) brings into visibility the property (\ref{eta}) for one chosen $\eta(f)$. The energy (\ref{Ener}) is also conserved but this property is not directly seen in (\ref{BEc}).

\subsection{Hamilton formulation of the reversible Boltzmann equation}\label{HamB}

The time evolution equations (\ref{BEpH}) are Hamilton's equations. The operator $L^{(p)}$, called a Poisson operator, can also be specified
by
writing the Poisson bracket
\begin{equation}\label{PBp}
\{A,B\}^{(p)}=(A_{\rr},A_{\vv})L^{(p)}\left(\begin{array}{cc}B_{\rr}\\B_{\vv}\end{array}\right)=A_{\rr}B_{\vv}-B_{\rr}A_{\vv}
\end{equation}
where $A$ and $B$ are sufficiently regular real valued functions of $(\rr,\vv)$. The bracket (\ref{PBp}) is indeed a Poisson bracket since it depends linearly on gradients of $A$ and $B$, the equality  $\{A,B\}^{(p)} =-\{B,A\}^{(p)}$ holds, and the Jacobi identity
$\{\{A,B\}^{(p)},C\}^{(p)}+\{\{B,C\}^{(p)},A\}^{(p)}+\{\{C,A\}^{(p)},B\}^{(p)}=0$ holds.

Since Eq.(\ref{BErev}) is just a Liouville equation corresponding to Eq.(\ref{BEpH}) and since (\ref{BEpH}) are Hamilton's equations, it is natural to ask the question as to whether (\ref{BErev}) is also Hamilton's equation. Indeed, Eq.(\ref{BErev}) is Hamilton's equation
\begin{eqnarray}\label{BEH}
\left(\frac{\partial f}{\partial t}\right)_{rev}&=&L^{(BE)}E_f\nonumber \\
&&=-\frac{\partial}{\partial\rr}\left(f\frac{\partial E_f}{\partial \vv}\right)+\frac{\partial}{\partial\vv}\left(f\frac{\partial E_f}{\partial \rr}\right)
\end{eqnarray}
where $E(f)$ is the energy given in (\ref{Ener});  $E_{f(\rr,\vv)}=\frac{\delta E}{\delta f(\rr,\vv)}$, $\frac{\delta \bullet}{\delta f(\rr,\vv)}$ is an appropriate functional derivative (in particular, for $E(f)$ given in (\ref{Ener}),  $E_{f(\rr,\vv)}=\frac{\vv^2}{2m}$).
Of course, we are not obliged to use in (\ref{BEH}) the energy (\ref{Ener}). If we use in (\ref{BEH}) a general energy (\ref{Energ})
where $e(f,\rr,\vv)$ is an arbitrary (but sufficiently regular) real valued function, we arrive at a general reversible Boltzmann equation. Hereafter, we shall consider  (\ref{BEH}) to be such equation.
The operator $L^{(BE)}$ appearing in (\ref{BEH}) is given by the Poisson bracket
\begin{equation}\label{PBBE}
\{A,B\}^{(BE)}=\int d\rr\int d\vv f(\rr,\vv)\{A_f,B_f\}^{(p)}
\end{equation}
where  $A$ and $B$ are sufficiently regular real valued function of $f$. The upper index $(BE)$ refers to Boltzmann Equation.
Straightforward verification shows that (\ref{BEH}) is indeed (\ref{BErev}) provided  $E^{(p)}(\rr,\pp)=\frac{\vv^2}{2m}$. With the Poisson bracket (\ref{PBBE}) we can write  the kinetic equation (\ref{BEH}) also as the equation
\begin{equation}\label{BEHP}
\dot{A}=\{A,E\}^{(BE)},\,\,for\,\,all\,\,A
\end{equation}
Indeed, $\dot{A}=\int d\rr\int d\vv A_f\frac{\partial f}{\partial t}$ and (by using integration by parts) $\{A,E\}^{(BE)}=\int d\rr\int d\vv A_f\left(-\frac{\partial }{\partial\rr}\left(f\frac{\partial E_f}{\partial \vv}\right)+\frac{\partial }{\partial\vv}\left(f\frac{\partial E_f}{\partial \rr}\right)\right)$.
Noncanonical Hamilton's equations in infinite dimensional spaces have started to be investigated by Clebsch in \cite{Clebsch}, later by Arnold in \cite{Arnold} and Marsden and Weinstein in \cite{MW} (where the proof that (\ref{PBBE}) is indeed a Poisson bracket can be found; note that the skew-symmetry of (\ref{PBBE}) is manifestly visible, only the Jacobi identity needs to be proved).

We turn  to
the property (\ref{eta}). In the context of the Hamiltonian formulation (\ref{BEH}) it   takes  the form
\begin{equation}\label{etaH}
\{A,S\}^{(BE)}=0\,\,for\,\,all\,\,A
\end{equation}
It can easily be shown  that $S(f)$ given in (\ref{Cas}) indeed verifies (\ref{etaH}). In the context of Hamilton's formulation, the entropy $S(f)$ arises as a Casimir function (the term used in the context of Hamilton's dynamics to call a function satisfying (\ref{etaH})). The function $\eta:\mathbb{R}\rightarrow\mathbb{R}$ introduced in (\ref{Cas}) will be  called a Casimir density. We note that the existence of Casimirs (that are different from a constant) for a bracket $\{,\}$   means that the bracket $\{,\}$ is degenerate. We also note that if we compare Godunov and Hamiltonian formulations  then the existence of a companion local conservation law in the context of Godunov formulation  means degeneracy of the Poisson bracket in the context of the Hamiltonian formulation

The Hamiltonian formulation (\ref{BEH}) has the following advantages.

(i)

The reversible Boltzmann equation (\ref{BErev}) is only one particular example of (\ref{BEH}) corresponding to one particular choice (\ref{Ener}) of the energy $E(f)$.

(ii)

The formulation (\ref{BEH})  displays manifestly its mechanical content. The vector field generating the time evolution (i.e. the right hand side of (\ref{BEH})) involves two separate elements:
kinematics (expressed in the operator $L$ or alternatively  in the Poisson bracket (\ref{PBBE})) and generating potential $E(f)$ that has the physical interpretation of energy. Only with the choice (\ref{Ener}) of the energy the general reversible Boltzmann equation (\ref{BEH}) becomes  the classical reversible Boltzmann equation (\ref{BErev}).

(iii)

The property (\ref{eta}) of solutions of the reversible Boltzmann equation appears as degeneracy (\ref{etaH}) of kinematics. From the physical point of view, we see that the conservation of the entropy in the reversible Boltzmann time evolution is a consequence of only kinematics. This property  has nothing to do with the specific choice of the energy $E(f)$. Contrary to the Godunov formulation (\ref{BEG}) where a specific function $\eta(f)$ is chosen at the outset of the analysis, we see in the Hamiltonian formulation the property (\ref{eta}) for any function $\eta:\mathbb{R}\rightarrow\mathbb{R}$.

\subsection{Irreversible Boltzmann equation}\label{IrrBE}

The reversible time evolution governed by (\ref{BErev})  preserves energy and entropy. Now we want to add to the right hand side of (\ref{BErev})  a term $\left(\frac{\partial f}{\partial t}\right)_{irrev}=\mathcal{B}(f)$ that is time irreversible (i.e. it does not change the sign if $\vv\rightarrow -\vv$) and that changes  solutions in such a way that the entropy  $S(f)$   growths   (i.e.  $\dot{S}(f) \geq 0$) while the energy  $E(f)$ remains  still  conserved. We are making this change in order to obtain a time evolution equation whose solutions will still agree with mechanics (i.e. the energy remains conserved) but they will also agree with the experimentally observed approach to equilibrium.
Following Boltzmann, we  see the physical origin of $\mathcal{B}(f)$ in the ignorance of the microscopic details of the particle trajectories undergoing binary collisions. We do not regard the binary collisions as mechanical events but as
"chemical reactions" $(\vv_1,\vv_2)\rightleftarrows (\vv'_1,\vv'_2)$ satisfying  constraints
\begin{equation}\label{const}
\vv_1+\vv_2=\vv'_1+\vv'_2;\,\,\,\vv_1^2+\vv_2^2=(\vv'_1)^2+(\vv'_2)^2
\end{equation}
and taking place at at one point with the coordinate $\rr_1$. The constraints (\ref{const}) expresses mathematically the mechanical origin of the "chemical reaction" $(\vv_1,\vv_2)\rightleftarrows (\vv'_1,\vv'_2)$, namely the momentum and the energy (only the kinetic energy) conservations. The remaining mechanical details (namely the time evolution equations (\ref{BEpH}) with the energy $E^{(p)}$ expressing the interaction between  the two colliding particles) are ignored. This ignorance is then the reason why the mechanics expressed in the Boltzmann equation is time irreversible and shows approach to equilibrium.

In order to obtain  the Boltzmann term $\mathcal{B}(f)$  we turn to
chemical kinetics (see \cite{MGchem}, \cite{MG84},\cite{Grrec}) and  write
\begin{equation}\label{BEir}
\left(\frac{\partial f(\rr_1,\vv_1)}{\partial t}\right)_{irrev}=\mathcal{B}(f)(\rr_1,\vv_1)=\Xi_{f^*(\rr_1,\vv_1)}
\end{equation}
where
\begin{equation}\label{Xi}
\Xi(X(f^*))=\int d1\int d1'\int d2\int d2'W(f,1,2,1',2')\left(e^{\frac{X}{2}}+e^{-\frac{X}{2}}-2\right)
\end{equation}
\begin{equation}\label{X}
X(1,2,1',2')=f^*(1)+f^*(2)-f^*(1')-f^*(2')
\end{equation}
and $W(f,1,2,1',2')\in\mathbb{R}^+$, that in chemical kinetics plays the role of rate coefficients,  is symmetric with respect to $1\leftrightarrows 2$ and $(1,2)\leftrightarrows (1',2')$, and different from zero only if the constraints (\ref{const}) are satisfied. We use shorthand notation $1=(\rr_1,\vv_1),...$.
The distribution functions $f^*$ are conjugate distribution functions introduced in (\ref{conj}).

The irreversible part $\mathcal{B}(f)$ of the vector field introduced above corresponds to the choice (\ref{Ener}) of the energy $E(f)$ and thus to the reversible part (\ref{BErev}) of the vector field. From the physical point of view, it expresses mathematically  the emergence of irreversibility due to our ignorance of details of  mechanics  of binary collisions. In the case of another choice of $E(f)$ (i.e. a choice different from (\ref{Ener})), there will be, in general, a different physical reason (i.e. different from the binary collisions) for the emergence of irreversibility and consequently different irreversible term $\left(\frac{\partial f}{\partial t}\right)_{irrev}$. Following \cite{GO}, we shall assume that from the mathematical point of view, $\left(\frac{\partial f}{\partial t}\right)_{irrev}$ will always the form (\ref{BEir}) but (\ref{Xi}), (\ref{X}) and the entropy will be different.

The complete kinetic equation governing the time evolution of $f$ is thus
\begin{eqnarray}\label{BEcomp}
\frac{\partial f}{\partial t}&=&
\left(\frac{\partial f}{\partial t}\right)_{rev} +\left(\frac{\partial f}{\partial t}\right)_{irrev}\nonumber \\
&&=L^{(BE)}E_f+[\Xi_{f^*}]_{f^*=S_f}\nonumber \\
&&=-\frac{\partial}{\partial\rr}\left(f\frac{\partial E_f}{\partial \vv}\right)+\frac{\partial}{\partial\vv}\left(f\frac{\partial E_f}{\partial \rr}\right)+[\Xi_{f^*}]_{f^*=S_f}
\end{eqnarray}

We make  five  observations about solutions to (\ref{BEcomp}).

\textit{Observation 1}

\begin{equation}\label{Hth}
\dot{S}=\int d1f^*(1)\Xi_{f^*(1)}
=\frac{1}{4}\int d1\int d2\int d1'\int d2'X\Xi_X\geq 0
\end{equation}

The proof is straightforward (see e.g. \cite{MGchem}). We recall that (\ref{eta}) implies $\dot{S}=0$ provided $\Xi=0$. We note that the inequality (\ref{Hth}) holds for any entropy $S(f)$ that is a Casimir of the Poisson bracket (\ref{PBBE}).

\textit{Observation 2}

The energy
$E(f)$ given in (\ref{Ener}) or in (\ref{Energ}) with $e(f,\rr,\vv)=\frac{v^2}{2m}+V_{pot}(\rr)$  (where $V_{pot}(\rr)$  is an unspecified potential energy $V_{pot}(\rr)$)    is conserved (i.e. Eq.(\ref{Edot}) holds). This result follows from (\ref{energ}) and from the constraints (\ref{const}).

\textit{Observation 3}

Eq.(\ref{BEir})  becomes the collision Boltzmann term  (see more in \cite{Grrec})
\begin{eqnarray}\label{Boltz}
\left(\frac{\partial f(\rr_1,\vv_1)}{\partial t}\right)_{irrev}&=&\int d2\int d1'\int d2'W^{(BE)}(f,1,2,1',2')\nonumber \\
&&\times(f(1')f(2')-f(1)f(2))
\end{eqnarray}
where
$W^{(BE)}(f,1,2,1',2')=\frac{2W(f,1,2,1',2')}{\sqrt{f(1)f(2)f(1')f(2')}}$ provided  the thermodynamic force (\ref{X}) is multiplied by $\frac{1}{k_B}$, where $k_B$ is the Boltzmann constant, and the entropy $S(f)$ is chosen to be  the Boltzmann entropy (\ref{Bent}), (\ref{Cas}).
Again, the proof is made by a direct calculation.
A closer investigation of  trajectories of particles undergoing binary collisions can relate $W^{(BE)}$ appearing in (\ref{Boltz}) (and also $W$ due to the relation between $W^{(BE)}$ and $W$ obtained above)  to the hard-core type repulsive potential among the particles.
In this paper we leave $W(f,1,2,1',2')$ specified only by the requirements, listed in the text following Eq.(\ref{X}), that guarantee  conservation of the kinetic energy and the momentum in binary collisions.

We note an important difference in the role that the entropy $S(f)$ plays in the reversible and irreversible part of the Boltzmann equation. In the reversible part it is a potential specified only by (\ref{Cas}). It means that if we restrict ourselves only to the reversible time evolution, we have no clue to choose a specific entropy. On the other hand, a specific choice of the entropy is a part of the formulation of the  irreversible part $\mathcal{B}(f)$ of the Boltzmann equation. We can write down $\mathcal{B}(f)$ either by choosing first the entropy and then writing it as (\ref{BEir}) or  by writing first  $\mathcal{B}(f)$ on the basis of some independent physical considerations (as Boltzmann did) and then,  by recasting  it into the form (\ref{BEir}), finding the entropy.

\textit{Observation 4}

As a consequence of (\ref{Hth}), (\ref{Edot}) and of another conservation law
\begin{equation}\label{Ndot}
\dot{N}=0
\end{equation}
where $N(f)$, having the physical interpretation of the number of moles, is given by $N(f)=\int d\rr\int d\vv f(\rr,\vv)$ [note that (\ref{Ndot}) is a straightforward consequence of (\ref{BErev}) and (\ref{const})],
solutions to (\ref{BEcomp}) approach, as $t\rightarrow \infty$, equilibrium distribution $f_{eq}(\rr,\vv)$ that is a solution of
\begin{equation}\label{Phif}
\Phi_f=0
\end{equation}
where
$\Phi(f,T,\mu)$, called thermodynamic potential, is given by
\begin{equation}\label{Phi}
\Phi(f,T,\mu)=-S(f)+\frac{1}{T}E(f)-\frac{\mu}{T}N(f)
\end{equation}
$T$ is the equilibrium temperature and $\mu$ the chemical potential, provided $\Phi$, as a function of $f$, is convex in a neighborhood of $f_{eq}(\rr,\vv)$.
The proof is based on the observation that the thermodynamic potential (\ref{Phi}) plays the role of the Lyapunov function corresponding to the approach to equilibrium.

Regarding the approach to $f_{eq}(\rr,\vv)$,
we note that the manifold of states satisfying $\dot{\Phi}=0$ (we shall denote it $\mathcal{M}_{irr}$) is much larger than $\mathcal{M}_{eq}$ (i.e. the manifold of equilibrium states $f_{eq}(\rr,\vv)$; $\mathcal{M}_{eq}\subset \mathcal{M}_{irr}$). The manifold $\mathcal{M}_{irr}$ is composed of local equilibrium (spatially inhomogeneous) states and the manifold  $\mathcal{M}_{eq}$ is composed of (spatially homogeneous) equilibrium states $f_{eq}(\rr,\vv)$. It has been shown in \cite{Grad},\cite{VillaniB} that, as $t\rightarrow \infty$,  solutions to the Boltzmann equation come near  $\mathcal{M}_{irr}$ but (due to the influence of the reversible part that by itself does not change the evolution of the thermodynamic potential $\Phi$) reach it only on $\mathcal{M}_{eq}$ where the time evolution stops.
We can also formulate the physical content of this important result about solutions of the Boltzmann equation as follows.  The origin of the dissipation that is explicitly present in the vector field of the Boltzmann equation is in microscopic details of particle trajectories and is, by itself, too weak to bring states to thermodynamic equilibrium states. However, due to the presence of the reversible part of the time evolution,  the weak microscopic dissipation trickles down to stronger and more macroscopic dissipation that brings eventually solutions to the Boltzmann equation to equilibrium states $f_{eq}(\rr,\vv)$.

\textit{Observation 5}

By evaluating the thermodynamic potential $\Phi$ at equilibrium states \\$f_{eq}(\rr,\vv,T,\mu)$ we obtain the equilibrium fundamental thermodynamic relation
\begin{equation}\label{eqsta}
[\Phi(f,T,\mu)]_{f=f_{eq}}=-\frac{PV}{T}
\end{equation}
where $P$ is the equilibrium pressure and $V$ is the volume of the space region in which the system under investigation is confined. It is easy to see (see e.g. \cite{GPK}) that if the energy $E(f)$ in (\ref{Phi}) is the energy (\ref{Ener}) and $S(f)$ is the Boltzmann entropy then the equilibrium fundamental thermodynamic relation (\ref{eqsta}) is the one corresponding to ideal gas. We thus see that the physical systems described by the classical Boltzmann equations cannot be more general than ideal gases. In the Hamiltonian formulation introduced in Section \ref{HamB} we can choose an arbitrary energy $E(f)$ and thus the domain of applicability becomes larger (e.g. Vlasov kinetic equation, describing a gas with an attractive long range interactions among the particles. is included). In all three formulations (i.e. the classical, the Hamiltonian, and the Godunov formulations) we can choose an arbitrary entropy of the type (\ref{Cas}).
But even with this freedom,  the domain of applicability of the Boltzmann equation (\ref{BEcomp}) is still very limited. For example, it can be shown (see \cite{vdW}) that in order that the fundamental thermodynamic relation  (\ref{eqsta}) changes from the one corresponding to an ideal gas to the one
corresponding to the van der Waals gas one needs not only to modify the energy (\ref{Ener}) by adding a long range attractive potential but also one needs to modify the entropy $S(f)$ by reaching outside the class of entropies defined in (\ref{Cas}). We expect to increase the domain of applicability of (\ref{BEcomp})  by making reductions discussed below.

We shall call hereafter the general kinetic  equation \eqref{BEcomp} a Kinetic Equation and we shall use the superscript $(KE)$ to denote the quantities related to it. The particular case of the Kinetic Equation is the Boltzmann equation \eqref{BEirr} and \eqref{Boltz}. The quantities related to the Boltzmann equation will be denoted by the superscript $(BE)$.

\section{Infinite Hamilton-Grad hierarchy}\label{GH}

Our objective now is to reduce the Kinetic Equation (\ref{BEcomp}). This means that we want to replace (\ref{BEcomp}) with simpler (hydrodynamics-type) time evolution equations whose solutions  will nevertheless reproduce (and moreover highlight) all the important features of solutions to (\ref{BEcomp}).

In order to formulate the reduction process we need the concept of the phase portrait. We recall that the phase  portrait $\mathcal{P}$ corresponding to the  time evolution equation generated by  the vector field $\mathcal{VF}$ is a collection of trajectories  for all possible initial conditions and a large class of material parameters through which the individual features of physical systems are expressed in $\mathcal{VF}$.

Reduction process consists of three steps: finding the phase portrait $\mathcal{P}^{(KE)}$ corresponding to the Kinetic Equation  vector field   $\mathcal{VF}^{(KE)}$,
recognizing in it a pattern $\mathcal{P}^{(red)}$, and finally identifying the vector field $\mathcal{VF}^{(red)}$ generating it. The vector field $\mathcal{VF}^{(red)}$ is then the vector field appearing in the
reduced, hydrodynamics-type,  time evolution equation. In other words, the reduction  \textit{kinetic  equation} $\rightarrow$ \textit{reduced equations} is a process
\begin{equation}\label{red}
\textit{Kinetic Equation} \rightarrow \mathcal{P}^{(KE)}\rightarrow \mathcal{P}^{(red)}\rightarrow \textit{reduced hydrodynamics-type eqs.}
\end{equation}
We note that the first arrow involves finding solutions to the Kinetic Equation  and the third arrow is in fact an inverse problem of solving the governing equations of reduced theories. In the third arrow we begin with  solutions and we look for the corresponding to it time evolution equations.

All three passages in (\ref{red})  are difficult to make and they all  can be made in many different ways. In this paper we  follow a novel route based on the Hamiltonian formulation. Its relation to more traditional approaches is discussed below in Section \ref{Compoth}.

To simplify the task of making the passages represented by the first and the third arrows (i.e. the task of solving the Kinetic Equation (\ref{BEcomp}) and the task of identifying the reduced time evolution equations from knowing their solutions), we simply express  the Kinetic Equation  vector field (i.e. the right hand side of (\ref{BEcomp})) in terms of quantities (called Grad's fields) that we can later interpret as state variables in the reduced theories. In other words, we replace trajectories with  infinitesimally short trajectories that are however expressed in a new form revealing new features.
In this viewpoint of the first and the third arrows we  follow Grad \cite{GradH} but with an important difference. We take as the point of departure the general kinetic equations (\ref{BEcomp}) (rather than (\ref{BErev}) together with (\ref{Boltz}) ) and insist on keeping  the Hamiltonian structure in all  reformulations. In fact, we  focus our attention in the reformulations only on  the Poisson bivector $L^{(KE)}$ (and the corresponding to it Poisson bracket (\ref{PBBE})) and  express it in terms of Grad's fields $\cc(\rr)=(c^{(0)}(\rr),...,c^{(\infty)}(\rr))$. Other parts of the structure of (\ref{BEcomp}), namely the energy and the irreversible vector field are left to be  unspecified  functions of Grad's fields. In this way we obtain, what we call, infinite Hamilton-Grad hierarchy. Its relation to the classical infinite  Grad hierarchy is discussed in Section \ref{Compoth}. The third arrow is in this approach completely absent  since the trajectories have been  replaced by vector fields. This means that after making the passage represented in (\ref{red}) by the second arrow we obtain the reduced time evolution equations that we search.

Regarding the second arrow in (\ref{red}) (discussed below in Section \ref{FGH}), we are suggesting that the pattern in which the extended hydrodynamics emerges  shows in the fields $\cc^{(\leq N)}=(c^{(0)},...,c^{(N)})$. This is because the Grad fields are chosen (see (\ref{moments}) below) in such a way that the first five of them,\\ $(c^{(0)}(\rr),c^{(1)}_{\alpha_1}(\rr),\frac{1}{2}c^{(2)}_{\alpha,\alpha}(\rr))$, represent the fields serving as state variables in the classical fluid mechanics. In order to simplify the notation we put hereafter the mass $m$ of one particle equal to one.  To  make the passage involved in the second arrow in (\ref{red}) means thus to find
realizations of the Hamilton-Grad  structure involving only a finite number of Grad's fields. This is a mathematical problem for which we do not have a general solution.
Except the case of five Grad fields (the case that corresponds to the classical fluid mechanics) the specific finite realizations that appear in  Section \ref{FGH} involve quantities that require to be specified by constitutive relations.  In this sense the governing equations  of extended hydrodynamics that arise in this paper in Section \ref{FGH} are similar to those arising  in other approaches. The equations themselves as well as the quantities that need constitutive relations for their specification    are however different due to our insistence on keeping  the Hamiltonian structure. For the same reason also the  methods that we are suggesting  to  use in  the investigations  of  constitutive relations are different from the methods suggested for the same purpose in other approaches.

In the rest of this section we want to reformulate (\ref{BEcomp}) into the form
\begin{equation}\label{HGcomp}
\frac{\partial\cc}{\partial t}=L^{(c)}E_{\cc}+[\Xi^{(c)}_{\cc^*}]_{\cc^*=S_{\cc}}
\end{equation}
We begin with the reformulation of the Poisson bracket (\ref{PBBE}) into a Poisson bracket $\{,\}^{(c)}$ involving Grad's fields $\cc(\rr)$. To make such reformulation, we do not need to express $f(\rr,\vv)$ in terms of $\cc(\rr)$. We need only a mapping from $f(\rr,\vv)$ to $\cc(\rr)$. Following Grad (see also \cite{MR}), we define this mapping by
\begin{eqnarray}\label{moments}
&& \cc=(c^{(0)},...,c^{(\infty)})\nonumber \\
&&c^{(i)}_{\alpha_1,...,\alpha_i}=\int d\vv v_{\alpha_1}...v_{\alpha_{i}}f(\rr,\vv); \,\,i=0,1,...,\infty;\,\,\alpha=1,2,3\nonumber \\
\end{eqnarray}
Let $M$ denote the space whose elements are one particle distribution functions $f(\rr,\vv)$. We can also see (\ref{moments}) as a way to  endow the space $M$ with a structure.

The mapping $f\mapsto \cc$ introduced in (\ref{moments}) implies the mapping $\cc^*\mapsto f^*$
\begin{equation}\label{fstar}
f^*(\rr,\vv)=\sum_{i=0}^{\infty}c^{(i)*}_{\alpha_1,...,\alpha_i}(\rr)v_{\alpha_1},...,v_{\alpha_i}
\end{equation}
where $\cc^*=S_{\cc}(\cc)$ and $f^*=S_f(f)$. This is because (\ref{moments}) implies
\begin{equation}\label{Af}
A_f=\sum_{i=0}^{\infty}A_{c^{(i)}_{\alpha_1,...,\alpha_i}}v_{\alpha_1}...v_{\alpha_{i}}
\end{equation}
where $A$ is a sufficiently regular function of $f$.

If we now replace $A_f$ and $B_f$ appearing
in (\ref{PBBE}) with (\ref{Af}), we arrive at (see also \cite{Ogul})
\begin{eqnarray}\label{PBGrad}
\{A,B\}^{(c)}&=&\int d\rr\sum_{i=0}^{\infty}\sum_{j=1}^{\infty}\sum_{k=1}^j c^{(i+j-1)}_{\alpha_1,...,\alpha_i,\beta_1,...,\beta_{k-1},\beta_{k+1},...,\beta_j}\nonumber \\
&&\times\left(\frac{\partial}{\partial r_{\beta_k}} \left(A_{c^{(i)}_{\alpha_1,...,\alpha_i}}\right)B_{c^{(j)}_{\beta_1,...,\beta_j}}
-\frac{\partial}{\partial r_{\beta_k}} \left(B_{c^{(i)}_{\alpha_1,...,\alpha_i}}\right)A_{c^{(j)}_{\beta_1,...,\beta_j}}\right)\nonumber \\
\end{eqnarray}

Leaving the energy $E=\int d\rr e(\cc(\rr))$   undetermined,  the Poisson bracket (\ref{PBGrad}) implies   the following  infinite hierarchy
\begin{eqnarray}\label{hierarc}
\left(\frac{\partial c^{(0)}}{\partial t}\right)_{rev}&=&-\sum_{j=1}^{\infty}\sum_{k=1}^j\frac{\partial}{\partial r_{\beta_k}}\left(c^{(j-1)}_{\beta_1,...,\beta_{k-1},\beta_{k+1},...,\beta_j}E_{c^{(j)}_{\beta_1,...,\beta_j}}\right)\nonumber \\
\left(\frac{\partial c^{(1)}_{\alpha}}{\partial t}\right)_{rev}&=&-\sum_{j=1}^{\infty}\sum_{k=1}^j\frac{\partial}{\partial r_{\beta_k}}\left(c^{(j)}_{\alpha,\beta_1,...,\beta_{k-1},\beta_{k+1},...,\beta_j}E_{c^{(j)}_{\beta_1,...,\beta_j}}\right)\nonumber \\
&&-\sum_{j=0}^{\infty}c^{(j)}_{\beta_1,...,\beta_j}\frac{\partial}{\partial r_{\alpha}} \left(E_{c^{(j)}_{\beta_1,...,\beta_j}}\right)\nonumber \\
&\vdots&\nonumber \\
\left(\frac{\partial c^{(N)}_{\alpha_1,...,\alpha_N}}{\partial t}\right)_{rev}&=&-\sum_{j=1}^{\infty}\sum_{k=1}^j\frac{\partial}{\partial r_{\beta_k}}\left(c^{(j+N-1)}_{\alpha_1,...,\alpha_N,\beta_1,...,\beta_{k-1},\beta_{k+1},...,\beta_j}E_{c^{(j)}_{\beta_1,...,\beta_j}}\right)\nonumber \\
&&-\sum_{j=0}^{\infty}\sum_{k=1}^{N}c^{(j+N-1)}_{\beta_1,...,\beta_j, \alpha_1,...,\alpha_{k-1},\alpha_{k+1},...,\alpha_N}\frac{\partial}{\partial r_{\alpha_k}} \left(E_{c^{(j)}_{\beta_1,...,\beta_j}}\right)\nonumber \\
\nonumber \\
&\vdots&
\end{eqnarray}

\subsection{Infinite Hamilton-Grad hierarchy with entropy}\label{HGE}

In the context of kinetic theory,  the entropy conservation is a consequence of the degeneracy (\ref{etaH}) of the Poisson bracket (\ref{PBBE}). In the context of the Hamiltonian-Grad hierarchy (\ref{hierarc}), we expect  to see it in the same way. In order to show it, we supplement  the moments $\cc$ given in (\ref{moments}) with  another moment
\begin{equation}\label{sb}
s(\rr)= b^{(0)}(\rr)= \int d\vv \eta(f(\rr,\vv))
\end{equation}
where $\eta(f)$ is the function introduced in (\ref{Cas}).   We are thus replacing $\cc$ given in (\ref{moments}) by
\begin{equation}\label{csdef}
(\cc,s)
\end{equation}
Since we call $S=\int d\rr\int d\vv \eta(f)$ entropy, we shall call the new moment $s(\rr)$ an entropy field.

With this extended set of moments we follow the steps that led us to (\ref{PBGrad}) and arrive (after lengthy calculations  - see \cite{Ogul} and Appendix in \cite{EntG} ) at
\begin{eqnarray}\label{PBcs}
\{A,B\}^{(cs)}&=&\{A,B\}^{(c)}\nonumber \\
&&+\int d\rr\sum_{i=1}^{\infty}\sum_{k=1}^{i}b^{(i-1)}_{\alpha_1,...,\alpha_{k-1},\alpha_{k+1},...\alpha_i}\nonumber \\
&&\times\left(\frac{\partial A_s}{\partial r_{\alpha_k}}B_{c^{(i)}_{\alpha_1,...,\alpha_i}}-\frac{\partial B_s}{\partial r_{\alpha_k}}A_{c^{(i)}_{\alpha_1,...,\alpha_i}}\right)
\end{eqnarray}
where
\begin{equation}\label{bmoments}
b^{(i)}_{\beta_1,...,\beta_j}=\int d\vv\eta(f)v_{\beta_1}...v_{\beta_j}; i=1,2,...
\end{equation}

The hierarchy corresponding to the bracket (\ref{PBcs}) has the form
\begin{eqnarray}\label{hihi}
\left(\frac{\partial s}{\partial t}\right)_{rev}&=&-\sum_{j=1}^{\infty}\frac{\partial }{\partial r_{\alpha}}\left(j b^{(j-1)}_{\beta_1,...,\beta_{j-1}}E_{c^{(j)}_{\alpha,\beta_1,...,\beta_{j-1}}}\right)\nonumber \\
\left(\frac{\partial c^{(0)}}{\partial t}\right)_{rev}&=&-\sum_{j=1}^{\infty}\sum_{k=1}^j\frac{\partial}{\partial r_{\beta_k}}\left(c^{(j-1)}_{\beta_1,...,\beta_{k-1},\beta_{k+1},...,\beta_j}E_{c^{(j)}_{\beta_1,...,\beta_j}}\right)\nonumber \\
&\vdots&\nonumber \\
\left(\frac{\partial c^{(N)}_{\alpha_1,...,\alpha_N}}{\partial t}\right)_{rev}&=&-\sum_{j=1}^{\infty}\sum_{k=1}^j\frac{\partial}{\partial r_{\beta_k}}\left(c^{(j+N-1)}_{\alpha_1,...,\alpha_N,\beta_1,...,\beta_{k-1},\beta_{k+1},...,\beta_j}E_{c^{(j)}_{\beta_1,...,\beta_j}}\right)\nonumber \\
&&-\sum_{j=0}^{\infty}\sum_{k=1}^{N}c^{(j+N-1)}_{\beta_1,...,\beta_j, \alpha_1,...,\alpha_{k-1},\alpha_{k+1},...,\alpha_N}\frac{\partial}{\partial r_{\alpha_k}} \left(E_{c^{(j)}_{\beta_1,...,\beta_j}}\right)\nonumber \\
&&-\sum_{k=1}^Nb^{(N-1)}_{\alpha_1,...,\alpha_{k-1},\alpha_{k+1},...,\alpha_N}\frac{\partial E_s}{\partial r_{\alpha_k}}\nonumber \\
&\vdots&
\end{eqnarray}
The energy $E$ appearing in (\ref{hihi}) is now a function of the fields (\ref{csdef}).

An obvious difference between the hierarchy (\ref{hierarc}) and the hierarchy (\ref{hihi}) is that (\ref{hihi}) involves fields $\bb=(b^{(1)},...,b^{(N)},...)$ that are not among the fields (\ref{csdef}) serving as state variables. However, the fields $\bb$ are related implicitly to $\cc$ by the map
$\cc\rightarrow \cc^*\rightarrow f^*\rightarrow f\rightarrow \bb$, where  $\cc^{*}(\cc)$  arises as a solution to $(-S^*(\cc^{*})+<\cc^{*},\cc>)_{\cc^{*}}=0$,
$S^*(\cc^*)=\int d\vv\int d\rr \eta^*(f^*(\cc^*))$, $<\cc^{*},\cc>=\int d\rr \int d\vv f^*(\rr,\vv)f(\rr,\vv)$ , and $f^*(\rr,\vv)=\sum_{i=0}^{\infty}c^{(i)*}_{\alpha_1,...,\alpha_i}(\rr)v_{\alpha_1},...,v_{\alpha_i}$.

\subsection{Properties of solutions of the infinite Hamilton-Grad hierarchy}\label{PS}

In this section we collect a few important properties of solutions of the Hamiltonian-Grad  hierarchy (\ref{hihi}).

\subsubsection{Mass conservation}

The second  equation in (\ref{hihi}) is in the form of the local conservation law
\begin{equation}\label{c0eq}
\left(\frac{\partial c^{(0)}}{\partial t}\right)_{rev}=-\frac{\partial J^{(0)}_{\alpha}}{\partial r_{\alpha}}
\end{equation}
where
\begin{equation}\label{J0}
J^{(0)}_{\alpha}=\sum_{j=1}^{\infty} j c^{(j-1)}_{\beta_1,...,\beta_{j-1}}E_{c^{(j)}_{\beta,\beta_1,...,\beta_{j-1}}}
\end{equation}
If we interpret physically  $c^{(0)}(\rr)$ as the mass density then Eq.(\ref{c0eq}) expresses the mass conservation and $\JJ^{(0)}$ is the mass flux. We note in particular that the mass flux is not only the momentum field (as it is in the classical Grad hierarchy  - see also Section \ref{10}) but it involves  all the higher order Grad fields on which the energy $E(\cc)$ depends.

\subsubsection{Entropy conservation}

Also the first equation in (\ref{hihi}) is the local conservation law
\begin{equation}\label{seq}
\left(\frac{\partial s}{\partial t}\right)_{rev}=-\frac{\partial J^{(s)}_{\alpha}}{\partial r_{\alpha}}
\end{equation}
where
\begin{equation}\label{Js}
J^{(s)}_{\alpha}=\sum_{j=1}^{\infty} j b^{(j-1)}_{\beta_1,...,\beta_{j-1}}E_{c^{(j)}_{\alpha,\beta_1,...,\beta_{j-1}}}
\end{equation}
is the entropy flux.

\subsubsection{Momentum conservation}\label{mass}

The third  equation in (\ref{hihi})
can also be cast into the local conservation law.
\begin{equation}\label{c1eq}
\left(\frac{\partial c^{(1)}_{\alpha}}{\partial t}\right)_{rev}=-\frac{\partial J^{(1)}_{\alpha,\beta}}{\partial r_{\beta}}
\end{equation}
where
\begin{eqnarray}\label{J1}
&&J^{(1)}_{\alpha\beta}=p\delta_{\alpha\beta}+\sigma_{\alpha\beta}\nonumber \\
&&\sigma_{\alpha\beta}=\sum_{j=1}^{\infty}j c^{(j)}_{\alpha,\beta_1,...,\beta_{j-1}}E_{c^{(j)}_{\beta,\beta_1,...,\beta_{j-1}}}\nonumber \\
&&p=-e+sE_s+\sum_{j=0}^{\infty}c^{(j)}_{\beta_1,...,\beta_j}E_{c^{(j)}_{\beta_1,...,\beta_j}}
\end{eqnarray}
We have used the relation
\begin{equation}\label{peqs}
s\frac{\partial}{\partial r_{\alpha}}E_s+\sum_{j=0}^{\infty}c^{(j)}_{\beta_1,...,\beta_j}\frac{\partial}{\partial r_{\alpha}} \left(E_{c^{(j)}_{\beta_1,...,\beta_j}}\right)=\frac{\partial p}{\partial r_{\alpha}}
\end{equation}
If we interpret $c^{(1)}$ as momentum density then (\ref{c1eq}) expresses the momentum conservation, $p$ is the scalar hydrodynamic pressure and $\ssigma$ is the stress tensor.

\subsubsection{Energy conservation}\label{ggen}

The energy $E(\cc)$  is conserved since the bracket $\{A,B\}^{(c)}$ given in (\ref{PBGrad}) is a Poisson bracket and $\dot{E}=\{E,E\}^{(c)}=0$. This equation, written as a local conservations, has the form
\begin{equation}\label{locen}
\left(\frac{\partial e}{\partial t}\right)_{rev}=-\frac{\partial J^{(e)}_{\alpha}}{\partial r_{\alpha}}
\end{equation}
where

\begin{eqnarray}\label{Je}
J^{(e)}_{\alpha}&=&(e+p)E_{c^{(1)}_{\alpha}}+\sigma_{\alpha i}E_{c^{(1)}_i}\nonumber \\
&&+\sum_{i=0}^{\infty}\sum_{j=2}^{\infty}\sum_{k=1}^j c^{(i+j-1)}_{\alpha_1,...,\alpha_i,\beta_1,...,\beta_{k-1},\beta_{k+1},...,\beta_j}E_{c^{(i)}_{\alpha_1,...,\alpha_i}}E_{c^{(j)}_{\beta_1,...,\beta_{k-1},\alpha,\beta_{k+1},...,\beta_j}}\nonumber \\
\end{eqnarray}

\subsubsection{Other  conservations}\label{otcon}

As we shall see below in Section \ref{Compoth},  the infinite Hamilton-Grad hierarchy (\ref{hierarc})   corresponding to the energy (\ref{Energ}), (\ref{Ener})   (i.e. (\ref{hierarc}) with   $E(\cc)=\int d\rr e(\cc(\rr))=\int d\rr \frac{1}{2}c^{(2)}_{\alpha \alpha}$) implies still other conservation laws (other than than those  seen above in this section). In fact, for this particular energy  the infinite Hamilton-Grad hierarchy  (\ref{hierarc}) implies an infinite number of conservation laws. All  equations  in the hierarchy are local conservation laws and thus
$\int d\rr \cc  $ are all conserved. This observation leads us to
ask the question of what are the properties that the energy $E(\cc)$ has to satisfy  in order that the hierarchy (\ref{hierarc}) implies
 an extra global conservation (i.e. extra to mass, energy, entropy, and momentum conservations that hold for all energies).
We shall answer this question for one particular extra global conservation, namely, we look for a condition on
the energy $E(\cc)$ in (\ref{hierarc}) that will guarantee that $\int d\rr c^{(2)}_{\alpha \alpha}$ be conserved in addition to the mass, energy,  and momentum.

From the third equation in (\ref{hierarc}) we see (by putting $N=2$) that
\begin{eqnarray}\label{c2}
\left(\frac{\partial c^{(2)}_{\alpha \alpha}}{\partial t}\right)_{rev}&=&-\sum_{j=1}^{\infty}\sum_{k=1}^j\frac{\partial}{\partial r_{\beta_k}}\left(c^{(j+1)}_{\alpha,\alpha, \beta_1,...,\beta_{k-1},\beta_{k+1},...,\beta_j}E_{c^{(j)}_{\beta_1,...,\beta_j}}\right)\nonumber \\
&&-2\sum_{j=0}^{\infty}c^{(j+1)}_{\alpha,\beta_1,...,\beta_j}\frac{\partial}{\partial r_{\alpha}} \left(E_{c^{(j)}_{\beta_1,...,\beta_j}}\right)\nonumber \\
\end{eqnarray}
This equation becomes a local conservation law if the second term on the right hand side becomes  divergence of a flux. We can write the second term as
\begin{equation}\label{third}
\frac{\partial}{\partial r_{\alpha}}\left(
-2\sum_{j=0}^{\infty}c^{(j+1)}_{\alpha,\beta_1,...,\beta_j}E_{c^{(j)}_{\beta_1,...,\beta_j}}\right) + 2\sum_{j=0}^{\infty}E_{c^{(j)}_{\beta_1,...,\beta_j}}\frac{\partial }{\partial r_{\alpha}}
c^{(j+1)}_{\alpha,\beta_1,...,\beta_j}
\end{equation}
Consequently, the equation that the energy $E(\cc)$ has to satisfy in order that the quantity $c^{(2)}_{\alpha\alpha}$ be conserved  is
\begin{equation}\label{eqc2}
\sum_{j=0}^{\infty}E_{c^{(j)}_{\beta_1,...,\beta_j}}\frac{\partial }{\partial r_{\alpha}}
c^{(j+1)}_{\alpha,\beta_1,...,\beta_j}=\frac{\partial J^{(2)}_{\alpha}}{\partial r_{\alpha}}
\end{equation}
where  $J^{(2)}$ is an unknown flux. The equation (\ref{eqc2}) is thus one equation for two unknown quantities $E(\cc)$ and $J^{(2)}(\cc)$.

We compare the analysis above with the analysis that we have already made in the discussion of the momentum conservation. If we rewrite the second term on the right hand side of the second equation in (\ref{hierarc}) in the same way as (\ref{third}), we arrive at the requirement
\begin{equation}\label{eqc3}
\sum_{j=0}^{\infty}E_{c^{(j)}_{\beta_1,...,\beta_j}}\frac{\partial }{\partial r_{\alpha}}
c^{(j)}_{\beta_1,...,\beta_j}=\frac{\partial J^{(1)}}{\partial r_{\alpha}}
\end{equation}
Obviously, this equation holds for $J^{(1)}=e$.

In general, we can formally regard the left hand sides of (\ref{eqc2}) and (\ref{eqc3}) as 1-forms $\xi=\sum_{j=0}^{\infty}E_{c^{(j)}}dc^{(j+1)}$ and $\zeta=\sum_{j=0}^{\infty}E_{c^{(j)}}dc^{(j)}$ respectively (the symbol $d$ denotes the exterior derivative). The requirements in Eqs.(\ref{eqc2}) and (\ref{eqc3}) are then requirements that  (if we use the terminology of differential geometry) the 1-forms $\xi$ and $\zeta$ are exact. The necessary condition for a form to be exact is that it is closed (i.e. $d\xi=0$ and $d\zeta=0$). We indeed see immediately that $d\zeta=0$ and  the 1-form $\zeta$ is closed. The equation $d\xi=0$ represents, if written explicitly, equations that the energy $E$, as a function of $\cc$,  has to satisfy in order that the form $\xi$ be closed.

\subsection{Infinite Grad hierarchy}\label{IH}

In Eq.(\ref{HGcomp}),  we have been so far investigation only the first term on its right hand side. The energy $E(\cc)$ (except in the discussion in Section \ref{otcon}) as well as the irreversible part of the vector field (i.e. the second term on the right hand side of (\ref{HGcomp})) have remained unspecified. We recall that since we are leaving open the specification of the energy $E(f)$  in the kinetic equation (\ref{BEcomp}), the term $[\Xi_{f^*}]_{f^*=S_f}$  generating the irreversible time evolution  has to be also left unspecified since $E(f)$  and $[\Xi_{f^*}]_{f^*=S_f}$ are related. We recall that
the Boltzmann collision term $\mathcal{B}$ expresses mathematically the physical insight according to which the ignorance of details of mechanics during
binary collisions is    the source of  irreversibility and dissipation. But this insight clearly applies only for the energy (\ref{Energ}), (\ref{Ener}) of a dilute ideal gas. For other energies,  other  insights are needed to identify the sources of dissipation.
We have assumed however that,  due to the requirement that the entropy increases in dissipation, the dissipative term in the kinetic equation has always the form $[\Xi_{f^*}]_{f^*=S_f}$ appearing in (\ref{BEcomp}). For the same reason we also require that the dissipative term in the Grad hierarchy will always have the form appearing in (\ref{HGcomp}). We shall discuss the choice of $E^{(c)}$ and $\Xi^{(c)}$ only in a few particular examples investigated in Section \ref{FGH} below.

\section{Examples of finite Hamilton-Grad hierarchies}\label{FGH}

Now we turn our attention to the second arrow in the reduction process (\ref{red}). Since the phase portrait $\mathcal{P}^{KE}$ became (due to our simplified discussion of the first arrow in (\ref{red})) just the vector field in the infinite Grd hierarchy $^{\infty}Gh$, and since  we physically interpret the lower  moments  as hydrodynamics or extended-hydrodynamics fields, the search for pattern in $\mathcal{P}^{KE}$ is the search for splitting   the infinite Grad hierarchy $^{\infty}Gh$ into \textit{reduced hierarchy} $^N[Gh]_{cl}$ that involves only the first $N$ Grad's fields (we denote them by the symbol $\cc^{(\leq N)}$) and the rest of the hierarchy,  $^{>N}Gh$,  called hereafter a \textit{reducing hierarchy}, that governs the time evolution of all Grad's fields of the order greater than $N$ (denoted $\cc^{(>N)}$). The reduced hierarchy $^N[Gh]_{cl}$ governs the time evolution of the fields $\cc^{(\leq N)}$ and does not involve any higher order fields. The time evolution equations constituting $^N[Gh]_{cl}$ are the governing  equations of the extended hydrodynamics that we search.

In the splitting of the infinite Grad hierarchy into reducing $^{>N}Gh$ and reduced $^N[Gh]_{cl}$ and then seeing in the latter the pattern that we search, we are making an assumption that the reducing time evolution (governed by $^{>N}Gh$ ) is faster than the reduced time evolution (governed by $^N[Gh]_{cl}$). Do  investigations of solutions to the kinetic equation (\ref{BEcomp}) support such assumption? For the ideal gas energy (\ref{Energ}), (\ref{Ener}) the answer is  "yes" for $N=5$ and  "no" for $N>5$ (see e.g. \cite{Weiss}, \cite{Santos}). From the physical point of view, this is a consequence of the fact that the Boltzmann collision operator $\mathcal{B}$ drives the distribution function to the local Maxwellian equilibrium (forming a submanifold $\mathcal{M}^{(lMaxw)}$ in the space of all distribution functions) that is parametrized by $\cc^{(\leq 5)}$. As rigorously proven in \cite{VillaniB}, during the time evolution governed by (\ref{BEcomp}) with  the ideal gas energy (\ref{Energ}), (\ref{Ener}),   the distribution functions come rapidly near to $\mathcal{M}^{(lMaxw)}$ and then, still remaining near to but never entering into  $\mathcal{M}^{(lMaxw)}$, continue to evolve towards the total equilibrium.  For this property we can call the manifold $\mathcal{M}^{(lMaxw)}$ a slow quasi-invariant manifold.
There is no "finer" structure in $\mathcal{B}$ that would allow the  existence of another quasi-invariant  manifold parametrixed by $\cc^{(\leq N)}$ for $N>5$. In the infinite Hamilton-Grad hierarchy (\ref{HGcomp}), this is indicated by the fact that the dissipative part of the time evolution is completely absent in the equations governing the time evolution of $\cc^{(\leq 5)}$ but present in all the remaining equations. This means that $\cc^{(\leq 5)}$ dissipate only indirectly through their coupling with higher order fields.

But the situation is completely different for the kinetic equation (\ref{BEcomp}) with a general energy $E(f)$ and a general (but compatible with the chosen energy $E(f)$) dissipation term. In the infinite Hamilton-Grad hierarchy (\ref{HGcomp}) we expect to see that the dissipation term becomes to be different from zero   only for large $N$. If this is the case then the assumption that the Grad fields with smaller $N$ evolve slower than the fields with larger $N$ may be justified.

Having made the assumption about the split into slow and fast time evolution, we now ask the question of how
we obtain $^N[Gh]_{cl}$ governing the slow time evolution.  We begin with the hierarchy $^NGh$, i.e. the hierarchy governing the time evolution of $\cc^{(\leq N)}$. This hierarchy   still involves however, at least in general, higher order Grad fields $\cc^{(>N)}$. The hierarchy  $^N[Gh]_{cl}$ is the closed hierarchy $^NGh$, i.e. the hierarchy $^NGh$ in which all the higher order fields $\cc^{(>N)}$ appearing in $^NGh$ are expressed in terms of $^{\leq N}\cc$. The relation
\begin{equation}\label{cll}
\cc^{(>N)}=\mathcal{C}l(\cc^{(\leq N)})
\end{equation}
is called a closure relation. In accordance with the terminology used in the classical hydrodynamics, we shall also call (\ref{cll}) a constitutive relation. We note that (\ref{cll}) defines a submanifold, we shall denote it by the symbol $\mathcal{M}_{cl}^{(N)}$, in the space whose elements are $(c^{(0)},...,c^{(\infty)})$.
How do we obtain the constituitive  relation (\ref{cll})? We shall discuss this question systematically later in Section \ref{crr}.

Below, we shall discuss in some detail  finite hierarchies with  $N=5$, $N=10$, and a finite hierarchy corresponding to the Cattaneo heat conduction theory.
In order  to simplify the notation, and also in order to bring the notation  closer to  hydrodynamics, we shall use the following symbols  for the first $13$ moments.
\begin{eqnarray}\label{hydnot}
c^{(0)}(\rr)=\rho(\rr)&=&\int d\vv f(\rr,\vv)\nonumber \\
(c^{(1)})_i(\rr)=u_i(\rr)&=&\int d\vv v_i f(\rr,\vv)\nonumber \\
(c^{(2)})_{ij}(\rr)=P_{ij}(\rr)&=&\int d\vv v_i v_j f(\rr,\vv)\nonumber \\
(c^{(3)})_{ikk}(\rr)=q_i(\rr)&=&\int d\vv v^2 v_i f(\rr,\vv)
\end{eqnarray}
We recall that (also for the sake of simplicity of the notation and  since we are putting our attention on the mathematical structures)  the mass $m$ of one particle has been  put equal to one.

\subsection{5-field Hamiltonian hydrodynamics}\label{5sec}

A remarkable feature of the Poisson bracket (\ref{PBcs}) is that if we limit the state variables $(\cc,s)$ only to $(c^{(0)},c^{(1)},s)$ (i.e. we limit the functions $A$ and $B$  in (\ref{PBcs}) only to those that depend only on  $(c^{(0)},c^{(1)},s)$ ) then (\ref{PBcs}) reduces to
\begin{eqnarray}\label{5}
\{A,B\}^{(5)}&=&\int d\rr \left[\rho\left(\frac{\partial}{\partial r_{\alpha}}(A_{\rho})B_{u_{\alpha}}-\frac{\partial}{\partial r_{\alpha}}(B_{\rho})A_{u_{\alpha}}\right)\right.\nonumber \\
&&\left.+s\left(\frac{\partial}{\partial r_{\alpha}}(A_{s})B_{u_{\alpha}}-\frac{\partial}{\partial r_{\alpha}}(B_{s})A_{u_{\alpha}}\right)\right.\nonumber \\
&&\left.+u_i\left(\frac{\partial}{\partial r_{\alpha}}(A_{u_i})B_{u_{\alpha}}-\frac{\partial}{\partial r_{\alpha}}(B_{u_i})A_{u_{\alpha}}\right)\right]
\end{eqnarray}
that does not involve  any other higher order field. The equations implied by (\ref{5})  are the familiar equations
\begin{eqnarray}\label{5eqs}
\left(\frac{\partial \rho}{\partial t}\right)^{(5)}_{rev}&=& - \frac{\partial J^{(5\rho)}_{\alpha}}{\partial r_{\alpha}}\nonumber \\
\left(\frac{\partial u_i}{\partial t}\right)^{(5)}_{rev}&=& - \frac{\partial J^{(5u)}_{i \alpha }}{\partial r_{\alpha}}          \nonumber \\
\left(\frac{\partial s}{\partial t}\right)^{(5)}_{rev}&=&- \frac{\partial J^{(5s)}_{\alpha}}{\partial r_{\alpha}}\nonumber \\
\left(\frac{\partial e}{\partial t}\right)^{(5)}_{rev}&=&- \frac{\partial J^{(5e)}_{\alpha}}{\partial r_{\alpha}}\nonumber \\
\end{eqnarray}
with
\begin{eqnarray}\label{5flux}
\JJ^{(5\rho)}&=&\rho E_{\uu}\nonumber \\
\JJ^{(5u)}&=&\ssigma^{(5)}+p^{(5)}\ddelta\nonumber \\
\JJ^{(5s)}&=&s E_{\uu}\nonumber \\
\JJ^{(5e)}&=&(e+p^{(5)})E_{\uu}
\end{eqnarray}
where
\begin{equation}\label{sig5}
\ssigma^{(5)}=\uu E_{\uu}
\end{equation}
and
\begin{equation}\label{pp5}
p^{(5)}=-e+\rho E_{\rho}+sE_s+<\uu, E_{\uu}>
\end{equation}
The last equation in (\ref{5eqs}) is a consequence (see Section \ref{ggen}) of the three equations above it but we have added it for completeness. Next, we explore some consequences of the fact that the bracket (\ref{5}) does not involve any other higher order field.

\textit{Comment 1}

We have seen above that the Hamiltonian kinematics of the 5-field nondissipative  hydrodynamics (expressed in the Poisson bracket (\ref{5})), splits out from the Hamiltonian kinematics of the infinite Grad hierarchy (expressed in the Poisson bracket (\ref{PBcs})) without any need for a closure or additional assumptions and simplifications. In addition, we see in (\ref{pp5}) that the 5-field Hamiltonian kinematics implies the local equilibrium (i.e. the quantity $p$ arising on the  right hand side the second equation in (\ref{5}) and having  the physical interpretation of pressure appears to be
related to the fields playing the role of state variables in the same way  as the thermodynamic pressure is related in the classical equilibrium thermodynamics to the thermodynamic mass and the thermodynamic entropy). This means that the Hamiltonian structure of the 5-field hydrodynamics implies the local equilibrium.

The Hamiltonian structure of the Euler hydrodynamics has been recognized first by Clebsch in \cite{Clebsch}. The Poisson bracket (\ref{5}), that has emerged in our analysis as a particular case of the general bracket (\ref{PBcs}), can also be derived by following at least two different and independent routes. On the first route \cite{Abram} it arises in the Lagrangian viewpoint of continuum (as an expression of kinematics of fluid particles).  A general association between the structure induced by a Lie group on the dual of its algebra and Poisson bracket \cite{MW} is the basis of the second derivation. The Lie group leading to (\ref{5}) is the group of   transformations $\mathbb{R}^3\rightarrow \mathbb{R}^3$ expressing kinematics of the continuum.

\textit{Comment 2}

Now we begin  with the Euler viewpoint  \cite{Euler} of the classical hydrodynamics  in which the governing equations (the Euler equations) arise as  an expression of Newton's law for continuum. The Euler equations constructed in this way turn out to be  Eqs.(\ref{5eqs}) with the energy $E$ that is a sum of the kinetic energy $\int d\rr \frac{\uu^2}{2\rho}$ and the internal energy that is independent of $\uu$. From the analysis presented above in this section we thus see that  the Euler hydrodynamics together with the assumption of local equilibrium  has the Hamiltonian structure. If we now combine this observation with the observation that we made in the previous comment, we see  that the 5-field Hamiltonian hydrodynamics is equivalent to the Euler hydrodynamics supplemented with the assumption of local equilibrium except for a difference in the energy $E$.
In the 5-field Hamiltonian hydrodynamics the choice of the energy $E$ is unrestricted,  in the Euler hydrodynamics the energy is a sum of the kinetic energy and the internal energy that is independent of the momentum field  (see more in Observation 4 in Section \ref{Compoth})).

\textit{Comment 3}

So far, we have been looking at   the passage from the infinite Hamilton-Grad hierarchy (\ref{hihi}) to the 5-field hydrodynamics (\ref{5eqs}) only from the point of view of the Hamiltonian kinematics (as its split  into the kinematics expressed in the Poisson bracket (\ref{5}) and the kinematics expressed in the Poisson bracket involving all the remaining terms in (\ref{PBcs})). Now, we  attempt to see it in the context of the reduction process (\ref{red}). This means that we want  to show that asymptotic solutions to the kinetic equation
can be well approximated by solutions to the governing equations of the 5-field hydrodynamics. Since we know that the Hamiltonian 5-field hydrodynamics implies the local equilibrium, solutions to the kinetic equation have to approach a distribution function expressing in the kinetic theory the local equilibrium. In the context of the ideal gas it is the local Maxwellian distribution. We can still continue and ask the question of what is the dissipative term in the kinetic equation that brings about the approach to the local Maxwellian distribution. The answer is well known, it is the Boltzmann collision term (\ref{BEir}). We therefore see that the Hamiltonian structure does not only imply the local equilibrium but also (indirectly) the binary collisions as  the source of the dissipation that brings about the approach (as $t\rightarrow\infty$) to the local Maxwellian distribution.

Summing up, we have shown that the Hamiltonian structure of the 5-field hydrodynamics is closely related to the local equilibrium and  (in the particular case of the ideal gas) to the Boltzmann collision term. This observation contributes to the understanding of the importance and the natural character of the assumption of local equilibrium in the classical hydrodynamics.

\textit{Comment 4}

In order to complete the 5-field hydrodynamics we still need to specify the energy $E(\cc^{(\leq 5)})$ and supply the reversible time evolution equations (\ref{5eqs}) with a dissipative term. As for the energy, if the energy with which we begin our analysis of $^{\infty}Gh$  depends on $\cc^{(> 5)}$ then we indeed need the closure (i.e. we need to investigate solutions of the reducing dynamics governed by $^{>5}Gh$). We also need to investigate solutions of $^{>5}Gh$ in order to bring the dissipation generated by the Boltzmann collision term to the dissipation of the hydrodynamic fields $\cc^{(\leq 5)}$. The archetype  analysis of this sort is the Chapman Enskog method.

Both the energy $E(\cc^{(\leq 5)})$ and the missing dissipation term in the Hamiltonian Euler hydrodynamic equations (\ref{5eqs}) can however be specified independently of  $^{\infty}Gh$ as an expression (on the level of description that uses the five hydrodynamic fields as state variables) of the physics of the particular fluid under consideration.
Regarding the choice of $E(\cc^{(\leq 5)})$, the physics involved is usually the fundamental relation of the thermodynamics of the particular fluid and the physics behind the choice of the dissipative term is the physics involved in the Navier-Stokes-Fourier constitutive relations. It has been shown (see e.g. \cite{Grrec},\cite{EntG}) that the Navier-Stokes-Fourier term in the hydrodynamic equations can be cast into the form of the third term on the right hand side of (\ref{HGcomp}).

\subsubsection{Godunov structure of the 5-field Hamiltonian hydrodynamics}\label{G5}

Another remarkable feature of the Hamiltonian Euler hydrodynamics (\ref{5eqs}) is that it possesses also the Godunov structure. Below, we shall recall the formulations revealing the structure.

The energy $E=\int d\rr e(\xx(\rr)) $ induces a conjugate field $e^*$ given by Legendre transform
\begin{equation}
\left(-e(\xx) + <\xx^*, \xx>\right)_{\xx} = 0,
\end{equation}
This equation specifies the relation $\xx(\xx^*)$.
The backward Legendre transformation,
\begin{equation}
\left(-e^*(\xx^*) + <\xx^*, \xx>\right)_{\xx^*} = 0,
\end{equation}
leads to the relation $\xx^*(\xx)$.

The formulation of the classical hydrodynamics displaying its Godunov structure is the following:
\begin{equation}\label{eq.CH.Godunov}
\left( \frac{\partial (e^*)_{ x^*_i}}{\partial t}\right)_{rev} = -\frac{\partial (J^*_k)_{x^*_i},}{\partial r_k},
\end{equation}
where the flux $\JJ^*$ is given by
\begin{equation}\label{eq.CH.Jk}
 J^*_k = e^* u^*_k.
\end{equation}
Left hand side of Eq.(\ref{eq.CH.Godunov}) is then simply equal to  $\dot{x}_i$.
Equation (\ref{eq.CH.Godunov}) can be  rewritten as
\begin{eqnarray}\label{5God}
  \left(\frac{\partial \rho}{\partial t}\right)_{rev} &=& -\frac{\partial}{\partial r_k} (\rho E_{u_k}),\\
  \left(\frac{\partial u_i}{\partial t}\right)_{rev} &=& -\frac{\partial}{\partial r_k} (e^*(\xx^*(\xx))\delta_{ki} + u_i E_{u_k})  \\
  \left(\frac{\partial s}{\partial t}\right)_{rev} &=& -\frac{\partial}{\partial r_k} (s E_{u_k}).
 \end{eqnarray}
These equations are exactly the same as the first three equations in (\ref{5eqs}).
Note that the scalar pressure arises now as conjugate of the entropy field.  This is indeed compatible with the expression (\ref{pp5}).

The Godunov Property 1 (see (\ref{BEc})) becomes now
\begin{eqnarray}
 \left(\frac{\partial e}{\partial t}\right)_{rev} &=& \left(\frac{\partial}{\partial t} \left(-e^* + x^*_i x_i\right)\right)_{rev}\nonumber\\
 &=& - (e^*)_{ x^*_i} \left(\frac{\partial x^*_i}{\partial t}\right)_{rev} + \left(\frac{\partial x^*_i}{\partial t}\right)_{rev} (e^*)_{ x^*_i} + x^*_i \left(\frac{\partial (e^*)_{x^*_{i}}}{\partial t}\right)_{rev}\nonumber\\
 &=& -x^*_i \frac{\partial (J^*_k)_{ x^*_i} }{\partial r^k} \nonumber\\
 &=&-\frac{\partial}{\partial r^k}\left(x^*_i (J^*_k)_{ x^*_i}\right) + \frac{\partial x^*_i}{\partial r^k} (J^*)_{x^*_i}\nonumber\\
 &=& -\frac{\partial}{\partial r^k}\underbrace{\left(-J^*_k + x^*_i (J^*_k)_{ x^*_i}\right)}_{J^{(e)}_k},
\end{eqnarray}
where $\JJ^{(e)}$ is the energy flux. This is then the fourth equation in (\ref{5eqs}).

The Godunov structure identified above is alternative to the structure identified in \cite{GR},\cite{GMR}.

\subsection{10-field Hamiltonian hydrodynamics}\label{10}

Next, we explore briefly extended hydrodynamics in which the tensor field $\PP$ (see (\ref{hydnot})) plays the role of an independent state variable. Altogether, we investigate thus hydrodynamics with 10 fields (two scalar fields $\rho$ and $s$, one vector field $\uu$ and one symmetric tensor field $\PP$). If we restrict the Poisson  bracket (\ref{PBcs})  to the functions $A$ and $B$ that depend only on the 10 fields $(c^{(0)},c^{(1)},c^{(2)},s)$ then we obtain (we use  the notation (\ref{hydnot}))
\begin{eqnarray}\label{10bracket}
\{A,B\}^{(10)}&=&\{A,B\}^{(5)}\nonumber \\
&&+\int d\rr
\left[u_l\left(\frac{\partial}{\partial r_{\alpha}}(A_{\rho})B_{P_{\alpha l}}-\frac{\partial}{\partial r_{\alpha}}(B_{\rho})A_{P_{\alpha l}}\right)\right.\nonumber \\
&&\left.+u_{\alpha}\left(\frac{\partial}{\partial r_{l}}(A_{\rho})B_{P_{\alpha l}}-\frac{\partial}{\partial r_{l}}(B_{\rho})A_{P_{\alpha l}}\right)\right.\nonumber \\
&&\left.+P_{i k}\left(\frac{\partial}{\partial r_{\alpha}}(A_{u_i})B_{P_{\alpha k}}-\frac{\partial}{\partial r_{\alpha}}(B_{u_i})A_{P_{\alpha k}}\right)\right.\nonumber \\
&&\left.+P_{i k}\left(\frac{\partial}{\partial r_{\alpha}}(A_{u_k})B_{P_{\alpha i}}-\frac{\partial}{\partial r_{\alpha}}(B_{u_k})A_{P_{\alpha i}}\right)\right.\nonumber \\
&&\left.+P_{i k}\left(\frac{\partial}{\partial r_{\alpha}}(A_{P_{i k}})B_{u_{\alpha}}-\frac{\partial}{\partial r_{\alpha}}(B_{P_{i k}})A_{u_{\alpha}}\right)\right.\nonumber \\
&&\left.+c^{(3)}_{i k l}\left(\frac{\partial}{\partial r_{\alpha}}(A_{P_{ik}})B_{P_{\alpha l}}-\frac{\partial}{\partial r_{\alpha}}(B_{P_{ik}})A_{P_{\alpha l}}\right)\right.\nonumber \\
&&\left.+c^{(3)}_{i k l}\left(\frac{\partial}{\partial r_{\alpha}}(A_{P_{il}})B_{P_{\alpha k}}-\frac{\partial}{\partial r_{\alpha}}(B_{P_{il}})A_{P_{\alpha k}}\right)\right.\nonumber \\
&&\left.+b^{(1)}_l\left(\frac{\partial}{\partial r_{\alpha}}(A_{s})B_{P_{\alpha,l}}-\frac{\partial}{\partial r_{\alpha}}(B_{s})A_{P_{\alpha l}}\right)\right.\nonumber \\
&&\left.+b^{(1)}_{\alpha}\left(\frac{\partial}{\partial r_{l}}(A_{s})B_{P_{\alpha l}}-\frac{\partial}{\partial r_{l}}(B_{s})A_{P_{\alpha l}}\right)\right]\nonumber \\
\end{eqnarray}
The time evolution equations corresponding to the bracket (\ref{10bracket}) are the following
\begin{eqnarray}\label{10eeqs}
\left(\frac{\partial \rho}{\partial t}\right)^{(10)}_{rev}&=& - \frac{\partial J^{(10,\rho)}_{\alpha}}{\partial r_{\alpha}}\nonumber \\
\left(\frac{\partial u_i}{\partial t}\right)^{(10)}_{rev}&=& - \frac{\partial J^{(10,u)}_{i \alpha }}{\partial r_{\alpha}}          \nonumber \\
\left(\frac{\partial s}{\partial t}\right)^{(10)}_{rev}&=&- \frac{\partial J^{(10,s)}_{\alpha}}{\partial r_{\alpha}}\nonumber \\
\left(\frac{\partial e}{\partial t}\right)_{rev}^{(10)}&=&- \frac{\partial J^{(10,e)}_{\alpha}}{\partial r_{\alpha}}\nonumber \\
\left(\frac{\partial P_{ik}}{\partial t}\right)^{(10)}_{rev}&=&
-\frac{\partial}{\partial r_{\alpha}}\left(P_{ik}E_{u_{\alpha}}+2c^{(3)}_{ikl}E_{P_{\alpha l}}\right)\nonumber \\
&&-2\overline{u_k\frac{\partial}{\partial r_i}(E_{\rho})}-2\overline{b^{(1)}_k\frac{\partial}{\partial r_i}(E_s)}\nonumber \\
&&-2\overline{P_{\alpha k}\frac{\partial}{\partial r_i}(E_{u_{\alpha}})}
-2\overline{c^{(3)}_{\alpha l k}\frac{\partial}{\partial r_i}(E_{P_{\alpha l}})}
\end{eqnarray}
with
\begin{eqnarray}\label{10flux}
\JJ^{(10,\rho)}&=&\JJ^{(5\rho)}+2\uu E_{\PP}\nonumber \\
\JJ^{(10,u)}&=&\ssigma^{(10)}+p^{(10)}\ddelta\nonumber \\
\JJ^{(10,s)}&=&\JJ^{(5s)}+2\bb^{(1)}E_{\PP}\nonumber \\
\JJ^{(10,e)}&=&(e+p^{(10)})E_{\uu}+\ssigma^{(10)} E_{\uu}+2E_{\rho}\uu E_{\PP}\nonumber \\
\end{eqnarray}
where
\begin{equation}\label{ssig10}
\ssigma^{(10)}= \ssigma^{(5)}+2\PP E_{\PP}
\end{equation}
and
\begin{equation}\label{pp10}
p^{(10)}=p^{(5)}+<\PP,E_{\PP}>
\end{equation}
By $\overline{\aaa}$ we denote symmetric part of the tensor $\aaa$.

We see that, contrary to the bracket $\{A,B\}^{(5)}$ obtained in (\ref{5}) by restricting the functions $A$ and $B$ in (\ref{PBcs}) only to functions that depend on $(c^{(0)},c^{(1)},s)$, the resulting bracket (\ref{10bracket}) involves two higher  order moments, namely   $c^{(3)}$ and $b^{(1)}$. The bracket (\ref{10bracket}) satisfies all the properties of the Poisson bracket (it is linear in gradients of $A$ and $B$, and $\{A,B\}^{(10)}=-\{B,A\}^{(10)}$) except the Jacobi identity $\{\{A,B\}^{(10)},C\}^{(10)}+\{\{B,C\}^{(10)},A\}^{(10)}+\{\{C,A\}^{(10)},B\}^{(10)}=0$ remains to be proved. We shall call such bracket pre-Poisson bracket and the mathematical structure that it represents a pre-Hamiltonian structure. The constitutive relations needed to specify the fields $c^{(3)}$ and $b^{(1)}$ appearing in (\ref{10bracket}) will be discussed in Section \ref{crr}.

Summing up, the governing equations   (\ref{10eeqs})  of 10-field nondissipative hydrodynamics represent a family of mesoscopic time evolution equations that possess the pre-Hamiltonian structure. Solutions to all equations in the family
conserve  mass, entropy, momentum, and energy. The mass flux, the entropy flux, the extra stress tensor,  and the scalar pressure are related to the ten  hydrodynamic fields $(\rho(\rr),s(\rr),\uu(\uu),\PP(\rr))$  by  (\ref{10flux}), (\ref{ssig10}),  and (\ref{pp10}).  The individual nature of the fluids under investigation is expressed in (\ref{10eeqs}) in the energy $E(\cc^{(<10)})$, in constitutive relations
for the symmetric tensor field $c^{(3)}(\rr)$, the vector field $ b^{(1)}(\rr)$,  and the irreversible part of the vector field. The ways to determine the constitutive relations will be discussed systematically in Section \ref{crr}.

\subsection{Cattaneo Hamiltonian hydrodynamics}\label{13}

In this illustrative example we turn to the Cattaneo \cite{Catt} extension of the Fourier description of heat conduction. The physical motivation for the extension is to transform  the Fourier heat transfer theory theory  into a new theory in which  the speed of the heat propagation is finite and  the domain of applicability extends to micro and nano scales \cite{Jou}.

The basic idea behind the Cattaneo extension is to adopt the heat flux as an extra independent state variable. We limit ourselves only to heat transfer in a medium that remains unchanged during the passage of heat. In the setting of the hierarchy (\ref{hihi}), we retain therefore only the fields $s(\rr),\qq(\rr))$  (we use the notation introduced in (\ref{hydnot}).

If we restrict the Poisson bracket (\ref{PBcs})  to functions $A$ and $B$ that depend only on  $(s(\rr),\qq(\rr))$ then we obtain
\begin{eqnarray}\label{Cattbr}
\{A,B\}^{(Catt)}&=&\int d\rr
\left[c^{(5)}_{\alpha l k }\left(\frac{\partial}{\partial r_{\alpha}}(A_{q_k})B_{q_l}-\frac{\partial}{\partial r_{\alpha}}(B_{q_k})A_{q_l}\right)\right.\nonumber \\
&&\left.+c^{(5)}_{\alpha l k }\left(\frac{\partial}{\partial r_{\alpha}}(A_{q_l})B_{q_k}-\frac{\partial}{\partial r_{\alpha}}(B_{q_l})A_{q_k}\right)\right.\nonumber \\
&&\left.+c^{(5)}_{ k }\left(\frac{\partial}{\partial r_{\alpha}}(A_{q_k})B_{q_{\alpha}}-\frac{\partial}{\partial r_{\alpha}}(B_{q_k})A_{q_{\alpha}}\right)\right.\nonumber \\
&&\left.+b^{(2)}_{\alpha  l}\left(\frac{\partial}{\partial r_{\alpha}}(A_{s})B_{q_{l}}-\frac{\partial}{\partial r_{\alpha}}(B_{s})A_{q_{l}}\right)\right.\nonumber \\
&&\left.+b^{(2)}_{\alpha  l}\left(\frac{\partial}{\partial r_{l}}(A_{s})B_{q_{\alpha}}-\frac{\partial}{\partial r_{l}}(B_{s})A_{q_{\alpha}}\right)\right.\nonumber \\
&&\left.+b^{(2)}\left(\frac{\partial}{\partial r_{\alpha}}(A_{s})B_{q_{\alpha}}-\frac{\partial}{\partial r_{\alpha}}(B_{s})A_{q_{\alpha}}\right)\right]\nonumber \\
\end{eqnarray}
The time evolution equations implied by (\ref{Cattbr}) are:
\begin{eqnarray}\label{Catteqs}
\left(\frac{\partial s}{\partial t}\right)^{(Catt)}_{rev}&=&-\frac{\partial J^{(Catt,s)}_{i}}{\partial r_i}\nonumber \\
\left(\frac{\partial e}{\partial t}\right)^{(Catt)}_{rev}&=&- \frac{\partial J^{(Catt,e)}_{i}}{\partial r_{i}}\nonumber \\
\left(\frac{\partial q_k}{\partial t}\right)^{(Catt)}_{rev}&=&-\frac{\partial}{\partial r_i}\left(c^{(5)}_{k}E_{q_i}\right)-2\frac{\partial}{\partial r_i}\left(c^{(5)}_{ilk}E_{q_l}\right)\nonumber\\
&&-b^{(2)}\frac{\partial E_s}{\partial r_k}-2b^{(2)}_{ik}\frac{\partial E_s}{\partial r_i}-c^{(5)}_{i}\frac{\partial E_{q_i}}{\partial r_k}-2c^{(5)}_{ilk}\frac{\partial E_{q_l}}{\partial r_i}
\end{eqnarray}
where
\begin{eqnarray}\label{fluxCatt}
J^{(Catt,s)}_{i}&=&b^{(2)}E_{q_i}+2b^{(2)}_{il}E_{q_l}\nonumber \\
J^{(Catt,e)}_{i}&=&b^{(2)}E_s E_{q_{i}}
+2b^{(2)}_{i k}E_s E_{q_k}
+2c^{(5)}_{i lk}E_{q_l}E_{q_k}+c^{(5)}_{k}E_{q_{i }}E_{q_k}
\end{eqnarray}

As we have already seen in the context of 10-field hydrodynamics, the bracket (\ref{Cattbr}) appears to be   unclosed, it involves  higher order fields $c^{(5)}$ and $b^{(2)}$ (we use in (\ref{Cattbr}) the notation:. $c^{(5)}_{ikl}=c^{(5)}_{iklmm}, c^{(5)}_{i}=c^{(5)}_{illmm}$, $b^{(2)}=b^{(2)}_{kk}$).  Again, as in the context of the bracket (\ref{10bracket}), we consider $c^{(5)}$ and $b^{(2)}$ as quantities that need to be specified in constitutive relations (discussed systematically in Section \ref{crr}).

Summing up, the time evolution equations (\ref{Catteqs}) posses the pre-Hamiltonian structure. Their solutions preserve the entropy and the energy. The energy and the entropy fluxes are related to the fields $(s(\rr),\qq(\rr))$ in  (\ref{fluxCatt}). What is missing in (\ref{Catteqs}) are the constitutive relations for the fields $c^{(5)}$ and $b^{(2)}$ and an appropriate dissipative term in the vector field. One example of the constitutive relations for $c^{(5)}$ and $b^{(2)}$  is presented below in
Section \ref{crr}. As for the dissipative part, we can follow the Cattaneo suggestion in which the dissipative term is proportional to $S_{\qq}$ or Guyer and Krumhansl \cite{GK} who have proposed the term involving also spatial gradients of  $S_{\qq}$ (see also \cite{GLD}).

\subsection{Constitutive relations}\label{crr}

The quantities that have remained undetermined in the extended hydrodynamic equations introduced above in this section need to be specified. Their specification is called, following the terminology established in fluid mechanics, constitutive relations. It is useful to put them into two classes. The energy, the entropy,  and the dissipation potential are specified in the constitutive relations belonging to the first class, the higher order fields appearing in brackets are specified in the constitutive relations belonging to the second class. Below, we shall  concentrate on the latter. The constitutive relations of the first class will be briefly mentioned in Section \ref{EXI}.

The bracket (\ref{PBcs}) does not need any closure, it is clearly a Poisson bracket. Also the bracket (\ref{5}) does not need a closure and is a Poisson bracket. All other brackets with the finite number $N$  of Grad's fields, obtained from (\ref{PBcs}) by restricting the functions $A$ and $B$ to those that depend only on $\cc^{(\leq N)}$,
(as e.g. (\ref{10bracket}) and (\ref{Cattbr}) ) involve higher order fields $\cc^{(>N)}$ that have to be specified by constitutive relations (\ref{cll}). How do we find the constitutive relations (\ref{cll})?  In general, we can follow   three routes described in the following three subsections.

\subsubsection{Jacobi identity}\label{JI}

The Jacobi identity $\{\{A,B\}^{(N)},C\}^{(N)}+\{\{B,C\}^{(N)},A\}^{(N)}+\{\{C,A\}^{(N)},B\}^{(N)}=0$  guarantees that the bracket $\{A,B\}^{(N)}$ is a Poisson bracket. We can see it
as an equation determining (or at least restricting the freedom of choice) of  the closure relation (\ref{cll}).
From the investigation reported in \cite{Ogul}, we note
that if we drop in (\ref{10bracket}) the last four lines then the bracket is closed and the time evolution equations (\ref{10eeqs}) with the kinetic energy playing the role of $E$ are identical with the 10-field hydrodynamics identified in \cite{Levermore}. The bracket (\ref{10bracket}) with the four last terms missing does not however satisfy the Jacobi identity.   But, the bracket (\ref{10bracket}) with the last five terms missing is the bracket that satisfies Jacobi identity and represents kinematics of the Reynolds  stress hydrodynamics (see \cite{Ogul}).

\subsubsection{Physical insight into kinematics}\label{phins}

We recall that the Poisson bracket $\{,\}^{(N)}$ is a mathematical expression of kinematics of the fields $\cc^{(\leq N)}$. This means that if we understand physically the kinematics we can find the Poisson bracket. Indeed, as we have discussed it in Comment 1 of Section \ref{5sec}, the Poisson bracket $\{,\}^{(5)}$ (that  arises in a  closed form, without any need of the closure, from the bracket (\ref{PBcs})) can  also be obtained from the physical understanding of the kinematics  of continuum (seen as the Lie group of the transformations $\mathbb{R}^3\rightarrow \mathbb{R}^3$). We shall now introduce a physical insight into the kinematics of the fields $(s(\rr),\qq(\rr))$ playing the role of state variables in the Cattaneo hydrodynamics in Section \ref{13} and arrive at constitutive relations for
$b^{(2)}_{ij}$ and  $c^{(5)}_{ijk}$ that  have remained  in the bracket (\ref{Cattbr}) undetermined.

It has been suggested in \cite{GT} to regard the Cattaneo hydrodynamics as the  two component hydrodynamics.  One component is the  material fluid with state variables $(\rho(\rr),\uu(\rr))$ and the other component is  the  caloric fluid with the state variables $(e(\rr),\qq(\rr))$. The same type of separation of the fields of extended hydrodynamics into material and caloric has appeared (apparently independently) in \cite{RS} (where the material fields are called F-fields and the caloric G-fields) and in thermal mass theories introduced in \cite{Huo}. On the basis of this insight we then  suggest that the kinematics of the fields $(s(\rr),\qq(\rr))$ is the same as the kinematics of the fields $(\rho(\rr),\uu(\rr))$. The Poisson bracket expressing mathematically the latter set of state variables is well  known (see Section \ref{5sec}) and thus  the Poisson bracket expressing the kinematics of $(s(\rr),\qq(\rr))$ is also known (see \cite{GLD} where such  bracket has been introduced).
What is the constitutive relation that transforms  the bracket (\ref{Cattbr}) into the bracket suggested in \cite{GLD}?
It can  easily be verified  that the  constitutive relation that makes such transformation is
\begin{eqnarray}\label{CC}
b^{(2)}_{ij}&=&\frac{1}{5} b s\delta_{ij}\nonumber \\
c^{(5)}_{ijk}&=&\frac{1}{15} c (q_i\delta_{jk}+q_j\delta_{ik}+q_k\delta_{ij})
\end{eqnarray}
where $b$ and $c$ are constants (material parameters) that have the physical dimension of energy. We are certain that with this constitutive relation the bracket (\ref{Cattbr})  satisfies the Jacobi identity. We are not however certain that the constitutive relation (\ref{CC})  is the only one for which (\ref{Cattbr})  has this property.

\subsubsection{Reducing dynamics}\label{reddyn}

We now return to (\ref{red}) and recall that the objective of reformulating   (\ref{BEcomp}) into (\ref{HGcomp}) is to recognize a pattern in $\mathcal{P}^{(KE)}$. Instead of looking at trajectories (i.e. solutions to (\ref{BEcomp})) we have been  looking  only at the infinitesimally short trajectories  (vector fields) that have been however    formulated in a new way that is expected to reveal the pattern. But even in the passage from the vector field (\ref{BEcomp}) to the vector field (\ref{HGcomp}),  we see now a need to  specify constitutive relations if we want to remain with only a finite number of Grad's fields. It may be useful to look for help in a deeper investigation of  solutions to  (\ref{BEcomp}).

So far, we have been concentrating our attention  only on  the reduced time evolution governed by $^N[Gh]_{cl}$. But the reducing (fast) time evolution governed by $^{>N}Gh$ provides  also a very interesting and very pertinent information. As it is argued in \cite{Gf}, the asymptotic solution of the reducing time evolution is in fact the closure relation needed to obtain the reduced (slow) time evolution. Moreover,  the reducing time  evolution introduces thermodynamics into the the fields $\cc^{\leq N)}$ (see \cite{Gf}).  Following this approach to constitutive relations,  we begin with the reducing hierarchy $^{>N}Gh$. We note  that the Poisson bracket that we need to construct its equations is directly seen in the bracket (\ref{PBGrad})). We simply restrict (\ref{PBGrad}) to functions $A$ and $B$ that are independent of $\cc^{(\leq N)}$).  The resulting bracket involves only the fields $\cc^{(>N)}$ and consequently there is no need for any closure. To specify the reducing time evolution is thus need only to   specify  the energy, the entropy, and the dissipation potential (all depending on the individual nature of the fluid under investigation and all  involving the fields $\cc^{(\leq N)}$). By solving the governing equations of the hierarchy $^{>N}Gh$ we arrive at the closure relation (\ref{cll}).
We shall return to this approach to constitutive relations  in Section \ref{Compoth}.

\subsubsection{Energy and dissipation potential}\label{EXI}

In the  passage from (\ref{BEcomp}) to (\ref{HGcomp}), we have discussed  so far only the Poisson bivector $L^{(c)}$. It remains to specify  the energy $E(\cc)$, the entropy $S(\cc)$,  and the dissipation potential $\Xi^{(c)}$. Contrary to $L^{(c)}$, the choice of these three quantities depends on the particular fluid under consideration.

For example, in the case of 5-field hydrodynamics, the specification usually proceeds as follows.  Either the energy field $e(\rr)$ ($E(\cc)=\int d\rr e(\rr)$) or the entropy field $s(\rr)$ ($S(\cc)=\int d\rr s(\rr)$) are one of the fields playing the role of state variables. Their mutual relation as well as  their  relation to $\uu$ and $\rho$ (that are the four remaining fields playing the role of state variables in the 5-field hydrodynamics) is different for different fluids. In the context of the classical hydrodynamics the relation  is  usually determined from the knowledge of thermodynamics  (i.e. from the knowledge of the fundamental thermodynamic relation representing the particular fluid under investigation in the cl;assical equilibrium thermodynamics)  of the fluid under consideration (by simply transposing this relation to the five hydrodynamic fields, or in other words, by  assuming local equilibrium). As for the dissipative part of the time evolution, the standard choice is the Navier-Stokes-Fourier dissipation. This dissipation term  is universal but it involves    several parameters (the viscosity and heat conductivity coefficients)  in which the individual nature of  the fluid under consideration is expressed.

In the case of N-field hydrodynamics, we need  a different insight for associating  $E(\cc)$,  $S(\cc)$, and  $\Xi^{(c)}$ with particular fluids.
In this paper we are not discussing particular fluids and   we  are therefore skipping  this part of constitutive relations.

\section{Relation to other approaches, further development}\label{Compoth}

As we have already mentioned in Introduction, investigations of Grad's hierarchies have contributed in an important way to the understanding of mesoscopic dynamics and thermodynamics. The investigation presented above in this paper continues this line of research. Our approach and our results are complementary to the approaches and to the results that can be found in the literature.
Below, we emphasize some of the features that distinguish our approach.
\\

\textbf{\textit{Top-down versus bottom-up approaches }}

The main difference between the approach followed in this paper and  the approach followed in the Extended Irreversible Thermodynamics \cite{Jou},  is in the starting point and in the spirit that is associated with it. The point of departure of \cite{Jou} is the classical hydrodynamics. The extension consists  in promoting the dissipative fluxes arising there into the status of independent state variables. We can regard this viewpoint as bottom up. The guiding principles in constructing the new time evolution equations are: the requirement of the compatibility with thermodynamics, the structure of the governing equations seen in the classical Grad hierarchy (see (\ref{Ghclass}) below), and the rich physical intuition originated in the classical hydrodynamics and in many specific examples of extensions. On the other hand, the point of view followed in this paper is top down. We begin with kinetic theory. Contrary to the approach that leads to the classical Grad hierarchy, we do not begin with a specific kinetic equation (as e.g. the Boltzmann equation) but with
the kinematics of kinetic theory. From this platform we then   look down on reduced descriptions. In the reductions, we  require  that the Hamiltonian structure of the reversible part of the time evolution that is seen in kinetic theory is preserved in the reduced mesoscopic theories.

The bottom up strategy of constructing
extended hydrodynamic equations has also been followed in the modeling of  flows of plastic
(viscoelastic) materials (see e.g. \cite{BirdI}), in particular then the flows  that arise during plastic processing operations in which melted plastic is flown into molds or is extruded. The usefulness of the resulting extended hydrodynamic theories  of complex fluids (i.e. fluids  involving an internal structure that evolves on  the  time scale that is comparable to the time scale on which the fluid as a whole evolves)
has played an important role in combating the sceptical view of extended hydrodynamics (based mainly on the results coming  from investigations of solutions to the  Boltzmann equation - we have  mentioned this type of  results in the second paragraph in Section \ref{FGH}).
\\

\textbf{\textit{Other kinetic equations serving as the point of departure of Grad's hierarchy}}

The starting point of our investigation of Grad's hierarchy is the Poisson bivector $L^{(BE)}$ appearing in the  family of general kinetic equation (\ref{BEcomp}). We have replaced the one particle distribution function $f(\rr,\vv)$ with Grad's fields $\cc(\rr)$ and transformed  the
Poisson bivector $L^{(BE)}$, expressing kinematics of $f(\rr,\vv)$, into the Poisson bivector $L^{(c)}$ expressing the kinematics of $\cc(\rr)$.
The energy, the entropy, and the dissipation potentials, depending on  the specific fluid under consideration, have remained undetermined.

It is, of course, possible to begin to investigate  the passage to the extended hydrodynamics   with other mesoscopic theories. Below, we shall briefly discuss the classical Grad hierarchy that unfolds  from the nondissipative Boltzmann equation (\ref{BErev}) and recall the passage to the extended hydrodynamics from several other mesoscopic theories.
\\

\textit{\textbf{Boltzmann's equation}}

The starting point of the classical investigation of the Grad hierarchy is the Boltzmann equation (i.e. Eq.(\ref{BErev}) supplemented with the Boltzmann collision term $\mathcal{B}$) that is a particular case of the Kinetic Equation (\ref{BEcomp}) corresponding to the particular choice  (\ref{Energ}), (\ref{Ener}) of the energy $E(f)$ and the particular choice  $\mathcal{B}$ of the dissipation term.

The Grad hierarchy corresponding to (\ref{BErev}) (i.e. the classical Grad hierarchy $^{\infty}Gh^{(class)}$) is simply obtained by
multiplying (\ref{BErev}) by $v_{\alpha_1}...v_{\alpha_{i}}$ and integrating over $\vv$.  We arrive at
\begin{eqnarray}\label{Ghclass}
\frac{\partial c^{(0)}}{\partial t}&=& -\frac{\partial c^{(1)}_{\alpha}}{\partial r_{\alpha}}\nonumber \\
\frac{\partial c^{(1)}_{\alpha_1}}{\partial t}&=& -\frac{\partial c^{(2)}_{\alpha_1,\alpha_2}}{\partial r_{\alpha_2}}\nonumber \\
&\vdots &\nonumber \\
\frac{\partial c^{(N)}_{\alpha_1,...,\alpha_N}}{\partial t}&=& -\frac{\partial c^{(N+1)}_{\alpha_1,...,\alpha_{N+1}}}{\partial r_{\alpha_{N+1}}}\nonumber \\
&\vdots &\nonumber \\
\end{eqnarray}
Regarding the dissipative part, we note that (due to the constraint (\ref{const})) the equations governing the time evolution of $c^{(0)},c^{(1)}$, and $tr c^{(2)}$ remain without the dissipative term.

We  now compare the classical Grad hierarchy (\ref{Ghclass}) with the Hamilton-Grad hierarchy (\ref{hierarc}) and make a few observations.

\textit{Observation 1}

We easily verify that the Hamilton-Grad  hierarchy (\ref{hierarc}) becomes,
with the energy $E(\cc)$  given by
\begin{equation}\label{EGrad}
E(\cc)=\int d\rr e(\cc(\rr))=\int d\rr \frac{1}{2}c^{(2)}_{\alpha \alpha}
\end{equation}
the classical  Grad hierarchy (\ref{Ghclass}). We therefore see that the relation between the classical hierarchy (\ref{Ghclass}) and the general hierarchy (\ref{hierarc}) is the same as the relation between  the Boltzmann equation and the Kinetic Equation (\ref{BEcomp}). The former is a particular case of the latter.
 This observation  brings immediately a new result about the classical hierarchy, namely that  \textit{the classical Grad hierarchy (\ref{Ghclass}) represents the Hamiltonian dynamics}.

\textit{Observation 2}

We note that all the time evolution equations in the hierarchy (\ref{Ghclass}) are local conservation laws (i.e. time derivative equals gradient of a flux) so that $\dot{C}^{(i)}=0$ for $i=1,2,...,\infty$, where $C^{(\alpha)}=\int d\rr c^{(\alpha)}$.

The existence of the companion conservation law representing physically the entropy conservation has been investigated previously in \cite{MR}, \cite{Levermore}. We shall only add to these investigations an observation that  the classical Grad hierarchy (\ref{Ghclass}) can indeed be cast into the Godunov form
\begin{equation}\label{Godc}
\left(\frac{\partial}{\partial t}(S^*)_{c^{(i)*}_{\alpha_1,...,\alpha_i}}\right)_{rev}=-\frac{\partial}{\partial r_k}(\JJ^*_k)_{c^{(i)*}_{\alpha_1,...,\alpha_i}}
\end{equation}
where
$S(\cc)$, called entropy,  is a real valued, sufficiently regular, and convex (or concave) function of $\cc(\rr)$, $ \cc^{(i)*}(\rr)=S_{\cc^{(i)}(\rr)}$,  $S^*(\cc^{*})$  is related to $S(\cc)$ by  $S^*(\cc^{*})=-S(\cc)+<\cc^{*},\cc(\cc^{*})>$, where $<\cc^{*},\cc>=\int d\rr \sum_{i=0}^{\infty}c^{(i)*}(\rr)c^{(i)}(\rr)$ and $\cc(\cc^{*})$ is a solution of $\left(-S(\cc)+<\cc^{*},\cc>\right)_{\cc}=0$. As in Section \ref{GodB},  we note that $\cc$ is conjugate to $S^*(\cc^{*})$ (i.e. $\cc^{(i)}(\rr)=S^*_{\cc^{(i)*}(\rr)}$) and also that  $S(\cc)=-S^*(\cc^{*}(\cc))+<\cc,\cc^{*}(\cc)>$ where
$\cc^{*}(\cc)$ is a solution of $\left(-S^*(\cc^{*})+<\cc^{*},\cc>\right)_{\cc^{*}}=0$. Moreover,
\begin{eqnarray}\label{L1}
S^*(\cc^*)&=& \int d\vv\int d\rr \eta^*(f^*(\cc^*))\nonumber \\
\JJ^*_k&=&\int d\vv v_k \eta^*(f^*(\cc^*))\nonumber \\
\end{eqnarray}
where the function $\eta(f)$ is the phase-space entropy density introduced in Section \ref{GodB} and $f^*$ is given by (\ref{fstar}).

\textit{Observation 3}

The fact that in the classical Grad hierarchy (\ref{Ghclass}) the higher order state variable is exactly the same as the lower order flux appearing in the previous time evolution equation is clearly  a consequence of
the specific (corresponding to the ideal gas) choice (\ref{EGrad}) of the energy. In the general Hamilton-Grad hierarchy (\ref{hierarc}),  the fluxes appearing in lower order equations are conjugates of   the higher order fields. The necessity to abandon the "rule" \textit{higher order state variable is the lower order flux} if passing from the  ideal gas to more general fluids has also been realized, on the basis of other types of considerations,  in \cite{Rugg}.

\textit{Observation 4}

The passage to the classical Euler hydrodynamics from the Hamilton-Grad hierarchy (\ref{hihi}) (see the three comments  in Section \ref{5sec} )is very different from the same type of passage   that starts in the  the classical hierarchy (\ref{Ghclass}).
In order to arrive at the Euler hydrodynamics from (\ref{Ghclass}), we proceed as follows. First, we keep in  (\ref{Ghclass}) only the equations governing the time evolution of $c^{(0)}$, $c^{(1)}$, and $tr c^{(2)}$. These equations involve higher order fields. We choose the standard constitutive relations that involve one scalar field $p(\rr)$. It is in the constitutive relations where we are stepping  beyond the ideal gas.
The  field $p(\rr)$ (having the physical interpretation of the local pressure) is then related to the fields  $c^{(0)}$, $c^{(1)}$, and $tr c^{(2)}$ by making the local equilibrium assumption. We thus see that the governing equations that appear directly and in a straightforward way (without making any assumptions) from the Hamilton-Grad hierarchy (\ref{hierarc}) appear from the classical hierarchy only after introducing constitutive relations and  assuming the  local equilibrium. But even after making these two assumptions, the resulting Euler hydrodynamics is still less general than the one arising from the Hamilton-Grad hierarchy (\ref{hierarc}). This is because the mass flux arising in (\ref{hierarc}) is the momentum field $\uu(\rr)$  only if the energy $E$ is the sum of the kinetic energy $\frac{\uu^2}{2\rho}$ and the remaining part that is independent of $\uu(\rr)$. In the fluids with strong spatial inhomogeneities (e.g. the fluids that are in the vicinity of gas-liquid phase transitions - see \cite{Grfl}) or certain complex fluids  - see \cite{Grmassfl}, \cite{VanG} the energy can depend also on the gradient $\nabla\uu $ of $\uu$ and consequently (see (\ref{5flux})) the mass flux   $E_{\uu}=\frac{\delta E}{\delta \uu}-\nabla\frac{\delta E}{\delta \nabla\uu}$  is  different from the classical mass flux $E_{\uu}=\frac{\delta E}{\delta \uu}=\frac{\delta \left(\int d\rr \frac{\uu^2}{2\rho}\right)}{\delta \uu}=\frac{\uu}{\rho}$  (see more in \cite{Grmassfl}, \cite{VanG}).

\textit{Observation 5}

Since the classical Grad hierarchy (\ref{Ghclass}) is a particular case of the Hamilton-Grad hierarchy (\ref{hierarc}), its solutions have all the properties proved  in Section \ref{PS} for the hierarchy (\ref{hierarc}). Due to the  specific nature and the relative (relative to (\ref{hierarc})) simplicity of the hierarchy (\ref{Ghclass}), many more detailed properties of its solutions  have been proven (see e.g.
\cite{Tr}, \cite{MR},\cite{Struch}, \cite{Weiss}, \cite{Santos}, \cite{K})
\\

\textit{\textbf{n-particle kinetic equations and kinetic equations with an internal structure}}

The most microscopic description of macroscopic systems is the description in which all position vectors and all momenta of all $\mathfrak{N}\sim 10^{23}$ microscopic particles composing the macroscopic system serve as state variables. Alternatively, the state variable could be  the $\mathfrak{N}$-distribution function $f_{\mathfrak{N}}(1,...,\mathfrak{N})$  or also the BBGKY sequence of distribution functions $(f_1(1),f_2(1,2),...,f_{\mathfrak{N}}(1,...,\mathfrak{N})$;  we are using   the short hand notation in which $n$ appearing in the set of the variables on which the distribution function depends means $(\rr_n,\vv_n)$. The equation governing the time evolution of the BBGKY sequence is the BBGKY hierarchy \cite{BBGKY}. This hierarchy is then
 the point of departure for the development of the extended hydrodynamics that  in the investigation reported in \cite{Kirkwood}. The Kirkwood approach  has been then  followed in rheology in \cite{BirdII} in the development  of the hydrodynamics of complex fluids like polymeric fluids and suspensions.
The Hamiltonian structures  have been introduced into this type of investigations in \cite{GPLA},  \cite{GPhD}, \cite{Beris}, \cite{Obook}. The generalized hydrodynamics arising in this analysis includes also two-point hydrodynamics \cite{GRII} in which the  fields serving as hydrodynamics state variables  depend on two position vectors. All the kinetic equations that have arisen in this type of extended kinetic theory can serve as the starting point for introducing new  Grad-type hierarchies.

Still another line of extensions of the Boltzmann kinetic theory has arisen by putting the dissipative part of the Boltzmann equation (i.e. binary collisions) into  focus. The strong interaction between the  particles undergoing  the binary collisions  brings  the internal structure of the particles into the analysis. In the investigation reported in \cite{Rugg},  the internal structure participating in the energy balance in binary  collisions is represented by one scalar  parameter $I\in [0,\infty)$. The one particle distribution function depends in this extended kinetic theory on $(\rr,\vv,I)$. The parameter $I$  appears then also in the definition of  Grad's fields. A very systematic investigation of Grad's hierarchy in this setting is presented in \cite{Rugg}.
\\

\textbf{\textit{Other mathematical structures}}

We have demonstrated in this paper the usefulness of the Hamiltonian structure in the development of extended hydrodynamics. There are, of course,   other mathematical structure that can also be very useful in the same type of investigation. Among the  structures that have    proven to play a very important role in extended hydrodynamics and that  are not included in the investigation presented  in this paper,  we mention three. First,  it is the invariance with respect to the  Galilean transformations. Consequences of the requirement of the Galilean invariance have been investigated very systematically in \cite{MR}, \cite{RGal}  and
more recently and in a somewhat different  mathematical setting in \cite{Van}, \cite{VanG}. The second structure is related to  the regularity of solutions, in particular then to the wave propagation and the  existence of shock waves (see \cite{MR},\cite{LiuRug}, \cite{Struch},\cite{Dafermos}, \cite{Weiss}). The third is the realizability (the question as to whether the finite set of the chosen moments  corresponds to a physically plausible, positive semi-definite distribution function $f(\rr,\vv)$ (see \cite{McD1},\cite{McD2}).
We intend to include these three aspects of solutions in our future investigations.
\\

\textbf{\textit{MaxEnt closure versus the closure obtained by solving reducing dynamics}}

We have suggested in Section \ref{reddyn} that one possible way to obtain   constitutive relations is  by constructing and solving  the reducing hierarchy $^{>N}Gh$. The constitutive relation becomes  the  asymptotic solution of $^{>N}Gh$. This is because we regard the time evolution governed by (\ref{BEcomp}) as proceeding in two stages: the fast (reducing) stage governed by $^{>N}Gh$ followed by the slow (reduced) stage governed by $\left[^NGh\right]_{\mathcal{C}l(\cc^{(\leq N)})}$, where $\mathcal{C}l(\cc^{(\leq N)})$ is the asymptotic solution of the fast dynamics. We now follow this viewpoint of the closure (or equivalently the viewpoint of the constitutive relations)   further and demonstrate  that  it becomes the same as the MaxEnt viewpoint  introduced in \cite{Dreyer} and followed in  \cite{Rugent}  provided  we make some simplifying assumptions about the reducing dynamics.

Let us assume that the reducing time evolution is governed by
\begin{equation}\label{Maxred}
\frac{\partial f}{\partial t}=-\Lambda^{(>N)} f S^{(>N)}_{f}(f)
\end{equation}
where $\Lambda^{(>N)}$ is an operator that is degenerate in the sense that $\Lambda^{(>N)}\vv_1...\vv_i=0$ for $i=0,1,...,N$ and positive definite in the rest of the space on which (\ref{Maxred}) is considered, $S^{(>N)}(f)$ is a sufficiently regular concave function of the distribution function $f$. The asymptotic solution $f^{(N)}_{ass}$ of (\ref{Maxred}) is the distribution function for which the potential
\begin{equation}\label{PhiN}
\Phi^{(>N)}(f;\bb)=-S^{(>N)}(f)+\sum_{i=0}^{N}b^{(i)}_{\alpha_1...\alpha_i}\int d\rr\int d\vv v^{(1)}_{\alpha_1}...v^{(i)}_{\alpha_i} f(\rr,\vv)
\end{equation}
reaches its minimum  The fields $\bb$ are the Lagrange multipliers that are related to the fields $\cc^{(\leq N)}$.

We note now that
if  $ S^{(>N)}(f)$ is the Boltzmann entropy (\ref{Cas}), (\ref{Bent}) then $f^{(N)}_{ass}$ obtained by looking for a minimum of (\ref{PhiN}) is the MaxEnt closure introduced in \cite{Dreyer}. We therefore see that if: (i) the reducing tome evolution is governed by (\ref{Maxred}), and (ii) the potential $S^{(>N)}(f)$ appearing in (\ref{Maxred}) is the Boltzmann entropy,  then the closure obtained by solving the reducing time evolution equation is the same as  the closure obtained by MaxEnt.  The simplifying assumptions  (i) and (ii) are however  very strong and  cannot be certainly applied universally. We make two observations.

If the kinetic equation serving as the point of departure is the Boltzmann equation and $N=5$ then (\ref{Maxred}) with $S^{(>5)}(f)$ being the Boltzmann entropy is  a good approximation of the reducing dynamics and the resulting $f^{(5)}_{ass}$  (that is in this case the local Maxwellian distribution) is the closure leading to the classical Euler hydrodynamics. Indeed,
$f^{(5)}_{ass}$ is the zero approximation  in the Chapman-Enskog solution of the Boltzmann equation.

If the starting kinetic equation is still the Boltzmann equation and $N>5$ then (due to the arguments presented in the second paragraph of Section \ref{FGH}) the time evolution equation (\ref{Maxred}) ceases to be a good approximation of the reducing dynamics. However, as we have already  discussed  also at the beginning of Section \ref{FGH}, if the kinetic equation serving as the point of departure is a general kinetic equation (\ref{BEcomp}) then (\ref{Maxred}) can still be a good approximation of the reducing dynamics. However, for a general kinetic equation,   the potential  $S^{(>N)}(f)$ will certainly be different from the Boltzmann entropy. A more detailed investigation of the reducing dynamics can be found  in \cite{Gf}.

\section{Concluding remarks}\label{Concl}

Mesoscopic time evolution is generated by a vector field that is a sum of two parts:  mechanical and thermodynamical.  The former is Hamiltonian, i.e. it is a gradient of energy transformed into a vector by a Poisson bivector expressing mathematically the kinematics of the state variable. The latter, which  makes its appearance  due to the ignorance of details of mechanics  brought about by replacing microscopic state variables with  mesoscopic ones, involves entropy and is compatible with the mechanical part in the sense that in the time evolution the energy is conserved and the entropy growths. This general framework, that we require to be kept in all formulations of the mesoscopic time evolution discussed in this paper,  guarantees by itself (i.e. irrespectively of the specific choice of the  undetermined  quantities in it) agreement with results of experimental observation of the approach of externally unforced fluids to equilibrium states at which the behavior appears to be well described by the classical equilibrium thermodynamics.

If the mesoscopic state variable is chosen to be the one particle distribution function then the mesoscopic theory is called a kinetic theory. With the special choice of the energy and the entropy corresponding to the ideal gas and choosing the binary collisions as the  source of dissipation, the kinetic equation (the equation governing the time evolution of the one particle distribution function) becomes the Boltzmann equation.

If the mesoscopic state variable is chosen to be the  infinite number of Grad's fields (that are velocity moments of the one particle distribution function) then the mesoscopic theory is called an infinite Grad's hierarchy. The Poisson bivector expressing mathematically kinematics of Grad's fields is derived in this paper rigorously from the Poisson bivector expressing kinematics in kinetic theory. The energy, the entropy and the dissipation term depend on the particularity of the fluid under consideration and their specification consequently requires additional physical insights. Contrary to the general infinite nondissipative Hamilton-Grad hierarchy that is introduced in this paper,  the classical infinite nondissipative  Grad hierarchy is only a straightforward reformulation of the nondissipative Boltzmann equation that  addresses, as well as the Boltzmann equation itself,  only the ideal gas. The general infinite nondissipative Hamilton-Grad hierarchy developed in this paper is therefore a more suitable  point of departure for investigating extended hydrodynamics.

The passage from the infinite hierarchy to a finite hierarchy (that can be physically interpreted as governing equations of extended hydrodynamics) is seen in this paper as splitting the infinite hierarchy into two hierarchies: one, called reduced hierarchy, is the finite hierarchy that we intend to interpret as representing the extended hydrodynamics, and the other hierarchy, called reducing hierarchy, is the remaining infinite hierarchy. Asymptotic solutions of the reducing hierarchy together with the requirement that the vector field generating the time evolution in the reduced hierarchy is again (as in the case of kinetic theory and the infinite Grad hierarchy) composed of the mechanical (Hamiltonian) part and the compatible with it thermodynamical  part then provide the constitutive relations needed in the reduced hierarchy.

In this paper we do not enter into a detail  investigation of specific reduced hierarchies representing extended hydrodynamics of specific complex fluids. The results of a general nature obtained in this paper have still however at least two  important specific implications.

The first one is about the fluxes that arise on the right hand side of the hydrodynamic equations. In the classical Grad hierarchy the  flux that arises in the equation governing the time evolution of the Grad field $c^{(N)}$ is the field $c^{(N+1)}$ . We see that this "rule" applies only for the ideal gas and is completely unusable, as we see in the general Hamilton-Grad hierarchy (\ref{hierarc}),  in the context of general fluids. In the general hierarchy, the right hand sides of the time evolution equations are not only divergences of fluxes,  and moreover, the fluxes that appear there are not the higher order fields but rather their conjugates.

The second implication  addresses  the role of the local equilibrium in the classical Euler hydrodynamics. We have shown that  the Euler hydrodynamic equations (i.e. nondissipative equations governing the time evolution of the fields of mass momentum and entropy) are Hamiltonian if and only if the fluid is in local equilibrium. This result contributes to the understanding of the importance and the natural character  of the local equilibrium assumption  in  the Euler hydrodynamics. Moreover, we emphasize that the Hamiltonian 5=field  hydrodynamics is still more general that the Euler hydrodynamics with the local equilibrium assumption since the former allows a mass flux that is different from the momentum field (see \cite{Grmassfl}, \cite{VanG}).
\\
\\
\\
\\

\textbf{Acknowledgements}

This research was partially supported by  the Natural Sciences and Engineering
Research Council of Canada. A part of the research was carried out in the Zhou Pei-Yuan Center for Applied Mathematics, Tsinghua University. One of the authors (MG) would like to thank Wen-An Yong, the members of the Center,  and  the students associated with it  for their  hospitality and stimulating discussions. Two authors (MG) and (GL) would like to acknowledge support received from the grant RI 15 biennum 2015-2017,  collaboration Bruxelles-Wallonie-Qu\'{e}bec. MP acknowledges support of Czech Science Foundation, project no.  17-15498Y.
\\
\\
\\

\end{document}